\documentclass{nature}

\usepackage[utf8]{inputenc}
\usepackage[english]{babel}

\usepackage[T1]{fontenc}
\usepackage{selinput}
\usepackage{babel}
\usepackage{ulem}

\usepackage{a4wide}
\usepackage{geometry}
\usepackage{hyperref}

\usepackage{color}
\usepackage{soul}

\usepackage{amsmath}
\usepackage{graphicx}
\usepackage{amssymb}

\usepackage{sidecap}
\usepackage{subfigure}

\usepackage{caption}
\providecommand{\aj}{Astron.\ J.}
\providecommand{\apj}{Astrophys.\ J.}
\providecommand{\apjl}{Astrophys.\ J.}
\providecommand{\apjs}{Astrophys.\ J.\ Suppl.}
\providecommand{\aap}{Astron.Astrophys.}
\providecommand{\physrep}{Phys.\ Rep.}
\providecommand{\nat}{Nature}

\providecommand{\prl}{Phys.\ Rev.\ Lett.}

\providecommand{\mnras}{Mon.\ Not.\ R.\ Astron.\ S.}
\providecommand{\apss}{Astrophys. Space Sci.}

\makeatletter % access to internal commands
\renewcommand\maketitle{\par
  \begingroup
    \if@twocolumn
      \ifnum \col@number=\@ne
        \@maketitle
      \else
        \twocolumn[\@maketitle]%
      \fi
    \else
      \newpage
      \global\@topnum\z@   % Prevents figures from going at top of page.
      \@maketitle
    \fi
    \thispagestyle{plain}\@thanks
  \endgroup
  \global\let\maketitle\relax
  \global\let\@maketitle\relax
  \global\let\@author\@empty
  \global\let\@date\@empty
  \global\let\@title\@empty
  \global\let\title\relax
  \global\let\author\relax
  \global\let\date\relax
  \global\let\and\relax
}

\usepackage[symbol]{footmisc}

\renewcommand\footnoterule{%
  \kern-3\p@\hrule\@width\columnwidth\kern 2.6\p@
}

\title{Quark deconfinement as supernova explosion engine for massive blue-supergiant stars}

\author{
Tobias~Fischer\footnote{~fischer@ift.uni.wroc.pl}$^{,1}$
Niels-Uwe~F.~Bastian$^1$,
Meng-Ru~Wu$^{2,3}$,
Petr~Baklanov$^{4,5}$,
Elena~Sorokina$^{4,6}$,
Sergei~Blinnikov$^{4,7}$,
Stefan~Typel$^{8,9}$,
Thomas~Kl{\"a}hn$^{10}$,
\&
David~B.~Blaschke$^{1,5,11}$
}

\begin{document}

\maketitle

\begin{abstract}
Blue-supergiant stars develop into core-collapse supernovae --- one of the most energetic outbursts in the universe --- when all nuclear burning fuel is exhausted in the stellar core. Previous attempts failed to explain observed explosions of such stars which have a zero-age main sequence mass of 50~M$_\odot$ or more. Here we exploit the largely uncertain state of matter at high density, and connect the modeling of such stellar explosions with a first-order phase transition from nuclear matter to the quark-gluon plasma. The resulting energetic supernova explosions can account for a large variety of lightcurves, from peculiar type II to super-luminous events. The remnants are neutron stars with quark matter core, known as hybrid stars, of about 2~M$_\odot$ at birth. A galactic event of this kind could be observable due to the release of a second neutrino burst. Its observation would confirm such a first-order phase transition at densities relevant for astrophysics. 
\end{abstract}
\begin{affiliations}
\tiny
\item Institute of Theoretical Physics, University of Wroc{\l}aw, Wroc{\l}aw, Poland
\item Institute of Physics, Academia Sinica, Taipei, Taiwan
\item Institute of Astronomy and Astrophysics, Academia Sinica, Taipei, Taiwan
\item National Research Center Kurchatov Institute, A.I. Alikhanov Institute of Theoretical and Experimental Physics, Moscow, Russia
\item National Research Nuclear University, Moscow, Russia
\item Sternberg Astronomical Institute, Moscow State University Moscow, Russia
\item Kavli Institute for the Physics and Mathematics of the Universe, University of Tokyo, Kashiwa, Chiba, Japan
\item Institut f{\"u}r Kernphysik, Technische Universit{\"a}t Darmstadt, Darmstadt, German
\item GSI, Helmholtzzentrum f{\"u}r Schwerionenforschung GmbH, Darmstadt, Darmstadt, Germany
\item Department of Physics and Astronomy, California State University Long Beach, California, USA
\item Bogoliubov Laboratory for Theoretical Physics, Dubna, Russia
\end{affiliations}

\newpage
Core-collapse supernova events concern stars with a zero-age main sequence (ZAMS) mass above $\sim$9~M$_\odot$. They are triggered by the  gravitational collapse of the stellar core. Associated explosions relate to the ejection of the stellar mantle\cite{Janka:2007}. The standard mechanism, due to neutrino heating, fails for very massive blue supergiant stars (BSGs) with a ZAMS-mass of 50~M$_\odot$ or more, some of which suffer multiple episodes of outbursts prior to the core-collapse event, associated with luminous blue variables (LBV)\cite{GalYam:2009,Foley:2011,Zhang:2012,Mauerhan:2013,Nicholl:2013,Terreran:2017}. Detailed numerical studies of such stellar progenitors yield the failed supernova explosion scenario, with the formation of black holes instead\cite{Sumiyoshi:2006id,Fischer:2009,OConnor:2011,Mueller:2017}. Besides a strong dependence on the progenitor star\cite{Woosley:2002zz,Umeda:2007wk}, one of the largest uncertainty is the high-density equation of state (EOS) of supernova matter, in particular at densities in excess of nuclear saturation density ($\rho_{\rm sat}=2.6\times10^{14}$~g~cm$^{-3}$ and $n_{\rm sat}=0.155$~fm$^{-3}$ in nuclear units) that cannot be probed by experiments. The only settled constraint is the high-precision observation of pulsars with masses of about 2~M$_\odot$\cite{Antoniadis:2013,Fonseca:2016}. Obeying this high-density constraint, we provide the connection between modeling the explosion of stars with ZAMS mass of 50~M$_\odot$ and the possibility of a first-order phase transition from ordinary matter, with baryonic (in general hadronic) degrees of freedom, to the quark-gluon plasma. This state of matter is expected to be encountered at high temperatures and densities. We model the entire supernova evolution using state-of-the-art general relativistic radiation-hydrodynamic simulations, with three-flavour Boltzmann neutrino transport\cite{Liebendoerfer:2004}. A new model EOS is developed within the two-phase approach featuring a first-order hadron-quark phase transition. We select a hadronic EOS\cite{Typel:2009sy} that is consistent with all current constraints\cite{Danielewicz:2002,Lattimer:2013,Krueger:2013,Antoniadis:2013,Abbott:2017}. For the quark-gluon plasma, first-principle calculations of Quantum Chromodynamics (QCD) -- the theory of strongly interacting matter -- are available only at vanishing density by means of solving the QCD equations numerically\cite{Laermann:2012PRL,Katz:2014PhLB}. They predict a cross-over hadron-quark transition at a pseudocritical temperature\cite{Bazavov:2011nk} of $T_c=154\pm9$~MeV (1~MeV~$\simeq 1.1\times 10^{10}$~K). Perturbative QCD is valid in the limit of asymptotic freedom, where quarks are no longer strongly coupled\cite{Kurkela:2014vha}. It automatically excludes astrophysical applications, since densities in the interior of neutron stars and supernovae are substantially below the asymptotic limit. Instead, effective quark-matter models are employed, e.g., the thermodynamic bag model\cite{Farhi:1984qu} and models of the Nambu-Jona-Lasinio (NJL) approach\cite{Nambu:1961tp}.
\\ \\
{\bf Quark matter at the supernova interior} \\
Hadron-quark EOS, which have been explored in core-collapse supernova studies so far\cite{Takahara:1988yd,Gentile:1993ma,Sagert:2008ka,Nakazato:2008su}, suffered from the lack of sufficient stiffness at high density and consequently low maximum masses. To overcome this deficiency a novel class of two flavor quark matter EOS has been developed\cite{Blaschke:2017}, based on the density-functional formalism. Deconfinement is taken approximately into account via an effective string-flip potential\cite{Horowitz:1985,Blaschke:1986}, henceforth denoted as SF. The inclusion of repulsive vector interactions among quarks provides stiffness with increasing density ($\rho\gg\rho_{\rm sat}$). This aspect is essential in order to yield massive hybrid stars $\gtrsim$2~M$_\odot$. SF includes linear vector interactions\cite{Klaehn:2015}, as in standard NJL-type models, and higher-order repulsive quark-quark interactions\cite{Benic:2014jia}. In order to employ SF in studies of core-collapse supernovae, we extend SF to finite temperatures and arbitrary isospin asymmetry, where little is known about the critical conditions for a possible hadron-quark phase transition in particular at high baryon density. It relates to the restoration of chiral symmetry while, on the other hand, the physical nature of (de)confinement is still unknown. The phase transition from nuclear matter\cite{Typel:2009sy} to the quark-gluon plasma is modeled via a Maxwell construction, resulting in an extended hadron-quark coexistence domain where the pressure slope is reduced compared to the two pure hadronic and quark phases, see Figs.~1a--1c. This has important consequences for the stability of compact stellar objects, which will be discussed further below. Note, only at zero-temperature and isospin symmetric matter, the pressure is exactly constant in the mixed phase (Fig.~1a). The SF model parameters are selected such that hybrid stars have a maximum mass of $2.17$~M$_\odot$ (with a radius of 10.2~km) and such that causality is fulfilled (the sound speed $c_s$ is strictly below the speed of light $c$). Note that the present parametrisation does not correspond to any of the original cases\cite{Blaschke:2017}, where $c_s \to c$. The resulting transition onset density of $3.5\times\rho_{\rm sat}$ for isospin symmetric matter at zero temperature is consistent with the constraint from the analysis of the elliptic flow from heavy-ion collisions\cite{Danielewicz:2002} (Fig.~1a). The phase boundaries vary only mildly with increasing isospin asymmetry, due to the inclusion of isovector-vector interactions ($\rho$-meson equivalent in hadronic matter) consistently in both phases. The corresponding neutron star configurations can be found in Supplementary Fig.~1 of the Supplementary Material. It includes mass-radius relation and corresponding EOS as well as tidal deformability. The latter agrees with the constraint deduced from the LIGO/Virgo collaboration analysis of the first binary neutron star merger GW170817\cite{Abbott:2017}, recently revisited independently\cite{Lattimer:2018}. With increasing temperature the hadron-quark phase boundary shifts to increasingly lower densities. This is illustrated for supernova matter (fixed baryonic charge-fraction of $Y_Q=0.3$ and constant entropy per particle of $S=3$~k$_{\rm B}$) in Figs.~1b and 1c. At low densities, $\rho\ll\rho_{\rm sat}$, this approach breaks down where the nature of the phase transition changes from first-order to smooth cross-over. However, the extrapolation of the hadron-quark phase boundary towards zero density yields a transition temperature of about $145$~MeV in agreement with $T_c$ from lattice-QCD\cite{Laermann:2012PRL,Katz:2014PhLB}.
\\ \\
{\bf Core-collapse supernova simulation} \\
We select a pre-collapse solar-metallicity stellar model with ZAMS mass of 50~M$_\odot$\cite{Umeda:2007wk}. It was evolved beyond the main sequence through all advanced nuclear burning stages. Several episodes of mass loss led to an extended envelop of $\sim$35~M$_\odot$ typical for BSG stars, with an enclosed mass of the nascent iron core of 1.89~M$_\odot$\cite{Umeda:2007wk}. We consider about 10$^5$~km of the stellar core, containing $\sim$9~M$_\odot$ of enclosed mass, including inner parts of the carbon-oxygen shell. This progenitor is evolved through stellar core-collapse and core bounce (the central density exceeds $\rho_{\rm sat}$ which defines the moment $t=0$), where the highly repulsive nuclear force balances gravity such that the stellar core collapse halts with the formation of a strong hydrodynamics shock wave\cite{Janka:2007} (black dashed line in Fig.~1d). The further evolution is determined by mass accretion onto the central compact proto-neutron star (PNS), from the gravitationally unstable layers above the stellar core. The entire evolution is illustrated in Fig.~1d, as low-entropy material ahead of the bounce shock composed of heavy nuclei falls onto the standing bounce shock where the entropy rises suddenly due to the dissociation of these nuclei. The PNS contracts accordingly (orange dash-dotted line in Fig.~1d), determined by the subsaturation density EOS. Until central densities of around $1.5\times\rho_{\rm sat}$ are reached, a few hundreds of milliseconds after core bounce, the evolution of the central temperature and density proceeds in a similar fashion for the massive ZAMS star of 50~M$_\odot$ and a stellar model with ZAMS mass of 12~M$_\odot$\cite{Woosley:2002zz} (Fig.~1c). The latter are canonically considered in core-collapse supernova studies. Differences arise during the later post-bounce evolution $\gtrsim500$~ms: strong mass accretion leads to massive PNSs which feature a fast mass-growth rate, and hence a steep gravitational potential. Significantly higher central densities and temperatures are reached for the more massive 50~M$_\odot$ ZAMS stellar model. This has important consequences. Up to now explosions of stars with ZAMS mass of $\sim$50~M$_\odot$ or more could not be obtained in supernova simulations that are based on microscopic models. The high compactness prevents the shock revival, e.g., via the delayed neutrino-heating scenario\cite{Bethe:1985ux} or magneto-rotationally driven explosions\cite{LeBlanc:1970kg,Bisnovatyi-Kogan:1976}. This still remains to be shown within self-consistent multi-dimensional supernova simulations\cite{Mueller:2017}. In all present models, the standing bounce shock retreats continuously. The enclosed mass inside the PNS grows until the maximum mass is reached, after which black hole forms (red dashed line in Fig.~1c) if no hadron-quark phase transition occurs\cite{Sumiyoshi:2006id,Fischer:2009,OConnor:2011}.
\\ \\
{\bf Hadron-quark phase transition} \\
This standard picture changes when a first-order hadron-quark phase transition is considered, for which the central density and temperature evolution is illustrated in Fig.~1c (thick black solid line). When the density exceeds the onset density for the phase transition, at $t=t_{\rm onset}\simeq1.110$~s, the PNS enters the hadron-quark mixed phase. The gradual softening of the EOS at the onset of the mixed phase (see Fig.~1b) results in a slow contraction of the PNS during which the mass enclosed inside the mixed phase grows for several tens to hundrets of milliseconds. At $t = t_{\rm collapse} = 1.215$~s, the critical conditions are reached (see Table~\ref{tab1}) and the PNS becomes gravitationally unstable. We define $t_{\rm collapse}$ as the moment when the interior PNS contraction velocities exceed few times $10^3$~km~s$^{-1}$. The contraction proceeds into a supersonic collapse on a timescale of less than one millisecond, towards the pure quark matter phase at higher density. The PNS collapse halts in the pure quark matter phase due to the increased stiffness at high density, aided by repulsive interactions, with the formation of an additional hydrodynamic shock wave. The release of latent heat drives the second shock wave from the PNS interior continuously towards larger radii. At the steep density slope at the PNS surface, the second shock accelerates and takes over the standing bounce shock, at a radius of  $\sim$80~km, which triggers the onset of the supernova explosion (Fig.~1d), at $t=t_{\rm final}=1.224$~s (henceforth denoted as $t_c$). The explosion shock expands quickly to radii of several 1000~km within less than 100~ms. Prior to the hadron-quark phase transition, the PNS mass grew continuously reaching about $M_{\rm PNS}=2.105$~M$_\odot$ of baryon mass (Fig.~2a). As a consequence of the phase transition, about 0.013~M$_\odot$ of hadronic matter is ejected from the PNS surface with the shock expansion (Fig.~2a). At the end of our simulations at about 10~s, the remnant contains 2.092~M$_\odot$ of baryon mass ($\sim$1.96~M$_\odot$ gravitational mass), with a massive quark-matter core of $M_{\rm quark}\simeq1.82$~M$_\odot$ (Fig.~2b), approaching the asymptotic value of $1.56$~M$_\odot$ at later times when the core temperature decreases and the PNS develops to the cold $\beta$-equilibrium remnant hybrid star. During the post-bounce evolution of the 12~M$_\odot$ progenitor, quark matter appears at $t_{\rm onset}=3.251$~s (thin blue dash-dotted line in Fig.~1c --- thick blue dash-dotted line: canonical post-bounce evolution up to  $\sim$1~s during which the hadron-quark phase boundary is not reached) if other mechanisms that revive the shock to trigger the explosion fail, e.g., neutrino heating\cite{Bruenn:2013,Mueller:2012a,Suwa:2013,Melson:2015}. Besides the longer delay, the supernova evolution associated with the hadron-quark phase transition proceeds qualitatively similar:  formation of a second shock wave which triggers of the explosion. We performed simulations for two more models with ZAMS masses of 18 and 25~M$_\odot$\cite{Woosley:2002zz}. All results are summarised in Table~\ref{tab1}. 
\\ \\
{\bf Prospects for detection} \\
The strong second shock features a rapid initial expansion with peak velocities on the order of the speed of light, which result in a high explosion energy of $E_{\rm expl}=3\times 10^{51}$~erg asymptotically (Fig.~2a). The explosion energy, mainly thermal energy, is converted into kinetic energy of the ejecta mass. It gives rise to average ejecta velocities of $\simeq3\times 10^3$~km~s$^{-1}$. In order to illustrate possible observational manifestations of the considered explosion mechanism, we first compute the composition of the ejected material via explosive nucleosynthesis calculations using an established nuclear reaction network\cite{Wu:2016}. During the initial shock expansion through the remaining inner parts of the silicon/carbon/oxygen layers, only a small amount of iron-group elements ($\simeq$0.024~M$_\odot$) is produced in comparison to canonical core-collapse supernovae ($\simeq$0.1~M$_\odot$). Observational properties of such an explosion in photons depend on the pre-collapse evolution, in particular on the mass-loss history. The latter aspect is largely uncertain. For the light curve modelling, we select two cases to demonstrate that supernova explosions of stars with ZAMS mass of 50~M$_\odot$, after undergoing the hadron-quark phase transition, can account for a wide variety of light curves. As the first case, we considered a BSG model\cite{Umeda:2007wk} where a small amount of mass has been lost. Almost the entire envelope ($\sim$35~M$_\odot$) remains as an atmosphere of the BSG at the moment of core collapse. Such 'compact envelope' results in a  dim light curve of type II peculiar supernova (Fig.~3), similar as SN1987A except the faster drop of the light curve here due to the smaller amount of iron-group elements produced. As second case, we assume a 'dilute envelope', which is related to the last stages of stellar evolution when the star can suffer abnormal mass loss as a result of unstable nuclear burning in the stellar bowels. If about half of the envelope mass is driven away few months prior stellar core collapse\cite{Zhang:2012}, and distributed within a radius $\sim$10$^{16}$~cm by the time of the SN explosion, the ejecta interaction with the circumstellar material is very effective to transform kinetic energy back into heating of the envelope which triggers the emission of photons. In this case we find a super-luminous supernova (SLSN) event, see Fig.~3, with a light curve similar to a hydrogen-rich SLSN such as the prototypical SN2006gy\cite{Smith:2007,Blinnikov:2012}. Furthermore, the rapid expansion of the second shock wave launched from the hadron-quark phase transition releases an observable neutrino signal: the propagation across the neutrinospheres launches a millisecond neutrino burst, few tens of milliseconds after $t_c$ (Figs.~4a and 4b). It is dominated by antineutrinos, unlike the $\nu_e$-burst associated with core bounce which is due to electron captures on protons. In contrast here, matter is dominated by neutrons and in the presence of a large abundance of positrons, due to the high temperatures achieved during the shock passage, neutrino-pair and $\bar\nu_e$-emission processes dominate over $\nu_e$. After that the neutrino luminosities and average energies decrease continuously, indicating the onset of supernova explosion, i.e. the beginning of the PNS deleptonization. Fig.~4c shows the number of expected neutrino events in the Super-Kamiokande detector for this supernova explosion at a distance of 10~kpc with 2~ms bins. Three extreme neutrino flavour oscillation scenarios are considered, including the case without oscillations as well as normal and inverted neutrino mass hierarchy, taking into account vacuum oscillations and the MSW-effect during the neutrino propagation through the stellar envelope. The blue crosses show the normal hierarchy case and the pink band illustrates the range sandwiched by these three scenarios. The additional effect due to non-linear collective flavor conversion\cite{Mirizzi:2016} should fall within this pink band. This clearly shows that the millisecond neutrino burst due to the phase transition will be detectable for a galactic event, independent of the detailed modeling of neutrino flavor oscillations. \\ \\
{\bf Summary and Discussion} \\
Self-consistent simulations of core-collapse supernovae are performed, launched from a massive star of 50~M$_\odot$ (ZAMS mass), during which the phase transition from ordinary hadronic matter to the quark-gluon plasma is considered. The latter relates to the releases of a large amount of latent heat and the formation of an additional supersonically expanding hydrodynamic shock wave, which launches the explosion. This is in contrast to the currently considered state-of-the-art failed supernova explosion scenario for such stellar progenitors, with the formation of black-holes instead. The small amount of nickel produced powers a dim light curve, if an ordinary BSG envelope of solar metallicity is present. The situation changes when the energetic ejecta collide with previously ejected material, e.g., due to winds and eruptions, which powers a bright light curve\cite{Blinnikov:2012} leading to super-luminous supernovae. The propagation of the second shock wave across the neutrinospheres releases an observable millisecond neutrino burst in all flavours, however, dominated by antineutrinos. This second neutrino burst contains characteristic information about the conditions of the hadron-quark phase transition, e.g., the delay of the second neutrino burst from the onset of neutrino detection is directly correlated with the conditions at the phase boundary, in particular the onset-density of the hadron-quark mixed phase. The width of the second neutrino burst is related to the energetics of the second shock, which depends on the latent heat. The quantitative behaviour may change in the presence of convection and hydrodynamics instabilities. In particular the delay of the second neutrino burst may be enhanced, which remains to be explored in multi-dimensional simulations. Nevertheless, due to the short timescale $\sim$1~ms of all processes associated with the hadron-quark phase transition -- PNS collapse, formation of the second shock, and its initial propagation -- it is unlikely that multi-dimensional phenomena may qualitatively alter the present findings. It may leave the possibility of the release of gravitational waves, due to the sudden rise of the central matter density associated with the high-density phase transition (Fig.~2c). \\
Critical for the launch of the second shock wave is the compactness of the stellar core, which is related to the enclosed mass inside the carbon-oxygen core. High compactness results in high mass-accretion rates during the post-bounce supernova evolution, leading to a rapid growth of the PNS mass. We performed additional long-term supernova simulations for stellar models with different ZAMS masses of 18 and 25~M$_\odot$\cite{Woosley:2002zz}. In the absence of an earlier explosion, all models undergo the hadron-quark phase transition. The corresponding conditions as well as explosion energy estimates are summarised in Table~\ref{tab1}, including the different times $t_{\rm onset}$: high(low) densities in the silicon layer above the iron core result in high(low) mass accretion rates during the post bounce evolution, consequently leading to a fast(slow) PNS contraction which drives the rate at which the central density and temperature rise (Supplementary Figure~2 illustrates this situation). Besides the different timing, all models follow the same evolution beyond the hadron-quark mixed phase: adiabatic collapse with supersonic velocities and the formation of a second shock wave, accelerating at the PNS surface propagating towards lower densities, which triggers the supernova explosion onset. Only in the case when the mass-growth rate exceeds the PNS compression, the enclosed mass will exceed the maximum mass given by the hybrid EOS (Table~\ref{tab1}), and the PNS will collapse to a black hole. This is the case for the 25~M$_\odot$ simulation (Supplementary Figure~2). Note that neutron stars with a mass below $\simeq1.56$~M$_\odot$ yield too low central densities to have a (pure) quark-matter core. They are most likely the remnants of supernova explosions of low- and intermediate-mass progenitors (ZAMS mass $\ll$50~M$_\odot$) for which the explosion is likely to start before reaching the hadron-quark mixed phase. Their evolution is then entirely determined by the hadronic EOS (a neutron star of $M=1.5$~M$_\odot$ has a radius of $\sim$12.2~km\cite{Typel:2009sy}). \\
Several of the findings reported here were observed before based on the bag model\cite{Dasgupta:2009yj}, being in conflict with several constraints such as too low maximum masses. A major step forward provided here is the QCD-motivated novel hadron-quark hybrid EOS for astrophysics, being in agreement with all presently known constraints. The inclusion of repulsive vector interactions ensures massive supernova remnant hybrid stars of about 2~M$_\odot$ at birth. The existence of such objects\cite{Tauris:2011,Tauris:2012,Antoniadis:2013,Fonseca:2016} is still a mystery. Simulations of 'canonical' core-collapse supernova explosions, of low- and intermediate-mass progenitors between 10--30~M$_\odot$ (ZAMS mass), yield remnant masses below $\sim$1.7~M$_\odot$\cite{Sukhbold:2016}. Alternatively, mass transfer in a binary system is not efficient enough due to low accretion rates $\sim10^{-9} - 10^{-6}$~M$_\odot$~yrs$^{-1}$ (for stable roche-lobe overflow) depending on the system's geometry\cite{Paczyski:1971}. Our scenario, if confirmed observationally, yields contributions to the high-mass tail around 2~M$_\odot$ of any possible neutron star distribution\cite{Ozel:2012,Antoniadis:2016}.
\\ \\ \\
{\Large Methods}
\smallskip
\\
{\small {\bf Core-collapse supernova simulations} are performed using the spherically-symmetric and fully general-relativistic radiation-hydrodynamics tool {\sc AGILE-BOLTZTRAN}\cite{Mezzacappa:1993gn,Liebendoerfer:2004}. It is based on three-flavour Boltzmann neutrino transport, with a complete set of all relevant weak processes, and flexible EOS module which was extended in this work with respect to the newly developed hadron-quark hybrid SF EOS. It features a conservative finite differencing scheme together with an adaptive-mesh refinement, both of which ensure accurate shock capture\cite{Liebendoerfer:2002}. We choose 208 radial mass-grid points and a neutrino distribution discretization of 24 energy and 6 scattering angle bins\cite{Mezzacappa:1993gn}. The explosion energy estimate, $E_{\rm expl}$, follows the common treatment; at every time-step the total specific energy -- the sum of internal, kinetic and gravitational specific energies -- is volume-integrated starting from the envelope towards the PNS surface. We also subtract the gravitational binding energy of the envelope. 
} \\ \\
{\small {\bf The SF EOS for quark matter} is based on the density-functional approach, depending on scalar and vector densities, $n_{\rm s}$ and $n_{\rm v}$, respectively\cite{Blaschke:2017}. It features deconfinement via an effective potential term $D_0\Phi(n_{\rm v})n_{\rm s}^{2/3}$, with vacuum string tension $D_0=(265~\rm MeV)^2$. The medium-dependent reduction of the string tension is modeled via the Gaussian functional $\Phi(n_{\rm v})=e^{-\alpha_0n_{\rm v}^2}$, with $\alpha_0 = 0.39$~fm$^{6}$. Repulsive vector interactions enter the density functional via the linear term, $a_0 n_{\rm v}^2$ with parameter $a_0=-4$~MeV~fm$^{3}$, and the higher-order term, $\gamma^{-1} b_0n_{\rm v}^4$ with parameter $b_0=1.6$~MeV~fm$^{9}$ and $\gamma = (1+c_0n_{\rm v}^2)$ with $c_0 = 0.025$~fm$^{6}$. The parameter $\alpha_0$ determines the onset density for deconfinement, while the parameters $b_0$ and $c_0$ are selected to ensure stable massive hybrid-star configuration of about 2.1~M$_\odot$, for matter in $\beta$-equilibrium at zero temperature. The $\gamma$-term is necessary to ensure causality, i.e. the sound speed never exceeds the speed of light. The extension of SF with respect to finite temperatures and arbitrary isospin asymmetry is straightforward since the original SF expressions are provided already in terms of Fermi-Dirac distribution functions where the temperature and the chemical potentials enter explicitly\cite{Blaschke:2017}. For the extension to arbitrary isospin asymmetry we include the isovector-vector term $\rho_\mathrm I \tau_3 n_\mathrm I^2$ in the density functional, i.e. the $\rho$ meson equivalent in hadronic matter. It depends on the isospin density $n_{\rm I}$ and where $\tau_{3}^{\rm up/down}$ denotes the third component of the isospin vector for up- and down-quarks. The additional parameter, $\rho_{\rm I}=80$~MeV~fm$^{3}$, is selected to ensure a smooth behaviour of the symmetry energy from the underlying hadronic EOS\cite{Typel:2009sy}. The latter aspect extends SF beyond the commonly employed approach of quark matter where the symmetry energy is due to the up- and down-quarks kinetic energy difference only. 
} \\ \\
{\small {\bf Neutrino oscillation and detection analysis} is performed by the same procedure described in detail in Sec.~VI of \cite{Wu:2015}. The assumed detector volume for supernova neutrino detection in Super-Kamiokande is taken as 32 kilotons\cite{Scholberg:2012}. The calculated neutrino detection events include all reaction channels: inverse-beta decay, neutrino-electron scattering, and neutrino-oxygen interactions (in bins of 2~ms), with tabulated interaction cross-sections\cite{Scholberg:2012} which assume the lowest interaction energy thresholds at 5~MeV. The neutrino energy spectra of different flavours without oscillations are taken directly from the supernova simulations. The effects of neutrino oscillations are modelled by adiabatic MSW three-flavour transformation with the measured vacuum mixing angles\cite{PDG}.
} \\ \\
{\small {\bf Lightcurve analysis} is performed employing the radiation-hydrodynamics package {\sc STELLA}, in its standard setup\cite{Blinnikov:2006,Sorokina:2016}. {\sc STELLA} employs a self-consistent treatment of hydrodynamics coupled with a multi-group radiative transfer, based on a variable Eddington-factor technique.} \\ \\
\\
\small Electronic supplementary material \\
\small {\bf Supplementary information} is available for this paper at https://doi.org/10.1038/s41550-018-0583-0.
\begin{addendum}

\item The authors are grateful for the discussion with K.-J.~Chen about possible implications regarding the supernova light curve, H.~Umeda for details into the stellar model used in this work, and A.~Yudin for helpful discussions regarding neutrino processes. The supernova simulations were performed at the Wroclaw Center for Scientific Computing and Networking (WCSS). The authors acknowledge support from the Polish National Science Center (NCN) under the grant numbers UMO-2016/23/B/ST2/00720~(TF and NUB) and DEC-2011/02/A/ST2/00306~(TF, NUB and DBB), the Russian Scientific Foundation (RSCF) under the grant numbers 16--12--10519 (PB and ES) and 18--12--00522 (SB), as well as from the Ministry of Science and Technology, Taiwan under Grant No. 107-2119-M-001-038 (MRW). DBB is further supported by the MEPhI Academic Excellence Project under contract No.~02.a03.21.0005 and ST acknowledges the DFG through grant SFB1245. This work was supported by the COST Actions CA15213 ``THOR'', CA16117 ``ChETEC'' and CA16214 "PHAROS".

\item[Author Contributions]
All the authors discussed the results and commented on the manuscript.
TF wrote the paper, implemented the hadron-quark EOS in the supernova model, performed all supernova simulations and analysed the corresponding results.
NUB, DB and TK developed the new quark EOS and the extension to finite temperatures and arbitrary isospin asymmetry.
ST provided the hadronic EOS selected for this study.
MRW performed the neutrino detection analysis and the nucleosynthesis calculations for the prediction of the elemental yields.
PB, ES and SB performed the lightcurve analysis.
All authors commented on the draft text.

\item[Data availability] The data that support the plots within this paper and other findings of this study, including the hadron-quark hybrid EOS, are available from the corresponding author upon request.

\item[\it{Competing Interests}] The authors declare that they have no competing financial interests.

\end{addendum}

\begin{table}
\centering
\caption{\small Summary of the SN simulation results with hadron-quark phase transition. The time $t_{\rm onset}$ refers to the moment when the PNS first reaches the hadron-quark mixed phase, while the labels ''collapse'' and ''final'' correspond to the onset of the the PNS collapse as well as when reaching the pure quark matter phase respectively. The latter is associated with the shock acceleration at the PNS surface. All times are relative to core bounce, $t-t_{\rm bounce}$, and the values for densities $\rho$ and temperatures $T$ are the central conditions. The explosion energy estimates, $E_{\rm expl}^*$, are associated with the moment of shock break-out, unlike the asymptotic value for the 50~M$_\odot$ simulation shown in Fig.~2.}
\begin{tabular}{ccccccccccc}
\hline
\hline
$M_{\rm ZAMS}$ & $t_{\rm onset}$ & $t_{\rm collapse}$ & $\rho\vert_{\rm collapse}$ & $T\vert_{\rm collapse}$ & $M^{1}_{\rm PNS, collapse}$ & $t_{\rm final}$ & $\rho\vert_{\rm final}$ & $T\vert_{\rm final}$ & $M^{1}_{\rm PNS, final}$ & $E_{\rm expl}^*$  \\
$[\rm M_\odot]$ & $[\rm s]$ & $[\rm s]$ & $[\rho_{\rm sat}]$& $[\rm MeV]$ & $[\rm M_\odot]$ & $[\rm s]$ & $[\rho_{\rm sat}]$& $[\rm MeV]$ & $[\rm M_\odot]$ & $[10^{51}~\rm erg]$ \\
\hline
12\cite{Woosley:2002zz} & 3.251 & 3.489 & 2.49 & 28 & 1.727 & 3.598 & 5.5 & 17 & 1.732 & 0.1 \\
18\cite{Woosley:2002zz} & 1.465 & 1.518 & 2.53 & 27 & 1.958 & 1.575 & 5.9 & 18 & 1.964 & 1.6 \\
25\cite{Woosley:2002zz} & 0.905 & 0.976 & 2.40 & 31 & 2.163 & 0.983 & 9.6 & 19 & 2.171$^2$ & -- -- \\
50\cite{Umeda:2007wk} & 1.110 &1.215 & 2.37 & 32 & 2.105 & 1.224 & 5.8 & 31 & 2.092 & 2.3 \\
\hline
\end{tabular}
\label{tab1} \\
{\small $^1$~baryon mass; $^2$~prompt black hole formation before shock breakout}
\end{table}

\newpage

{\Large \bf \hspace{-12mm} Figure legends} \\ \\
{\bf Figure~1.}~$\vert$ {\bf Hybrid EOS, phase diagram and supernova evolution.} {\bf a} and {\bf b.} pressure $P$ with respect to density $\rho$ in terms of $\rho_{\rm sat}$ for the SF hadron-quark hybrid EOS, for isospin symmetric matter at zero temperature, in agreement with the elliptic flow data from heavy-ion collisions\cite{Danielewicz:2002} {\bf (a)}, and for supernova matter (fixed baryonic charge-fraction $Y_Q=0.3$ and entropy per particle of $S=3$~k$_{\rm B}$) {\bf (b)}. The grey shaded areas show the density domain of the hadron–quark mixed phase. {\bf c.} temperature-dependence (in units of the pseudo-critical temperature $T_{\rm c}$) of the phase boundaries for the onset density of the hadron-quark phase transition (dark grey dashed line) and the pure quark matter phase (dark grey dash-dotted line). The maximum density scale corresponds to $6\times\rho_{\rm sat}$. Curves of constant entropy per particle (green solid lines) in units of $k_{\rm B}$ extend through the hadron-quark mixed phase (grey shaded region). Supernova evolution trajectories are shown, corresponding to the central conditions, for the 50~M$_\odot$ ZAMS star considered in this work (thick black solid line: with hadron-quark phase transition which triggers the supernova explosion; red dashed line: black-hole formation due to the failed supernova explosion without hadron-quark phase transition) in comparison to the evolution of a 12~M$_\odot$ ZAMS star\cite{Woosley:2002zz} (thick blue dashed line: standard supernova evolution up to about 1~s post bounce during which the hadron-quark phase boundary is not reached; thin blue dashed line: SN simulation continued until about 10~s after bounce during which the hadron-quark phase transition occurs at $t_c\simeq 3.598$~s). The maximum central densities shown here (thick solid black line and thin blue dash-dotted line) are reached right after the PNS collapse halts and the second shock wave forms. The density oscillations and the later evolution as shown in Fig.~2c are not included. {\bf d.}~space-time diagram of the evolution corresponding to the supernova simulation of the 50~M$_\odot$ progenitor with hadron-quark phase transition at $t_{\rm c}=1.224$~s. The evolution of the bounce shock (black dashed line) and the PNS radius (orange dash-dotted line) are marked together with the rise of the second shock wave (solid black line) which appears due to the hadron-quark phase transition.
\\ \\ \\ \\
{\bf Figure~2.}~$\vert$ {\bf Evolution of the supernova simulation} for the 50~M$_\odot$ ZAMS star based on the hadron-quark hybrid EOS where the phase transition takes place at $t_c$, shown in Fig.~1d. {\bf a} and {\bf b.} explosion energy $E_{\rm expl}$ as well as enclosed baryon mass inside the PNS, $M_{\rm PNS}$, and mass enclosed inside the quark-matter core, $M_{\rm quark}$. {\bf c.} central rest-mass density, $\rho_{\rm central}$. \\ \\
{\bf Figure.~3.}~$\vert$ {\bf Bolometric light curve} of the supernova simulation launched from the 50~M$_\odot$ ZAMS star with the hadron-quark phase transition. Two different density distributions of the 35~M$_\odot$ stellar envelope are considered, a compact envelope with no mass loss leading to the peculiar SN1987A-like event (solid line) and a dilute envelope with extensive mass loss giving rise to SLSN (dashed line). \\ \\
{\bf Figure~4.}~$\vert$ {\bf Neutrinos} from the supernova simulation of the 50~M$_\odot$ ZAMS star with the hadron-quark phase transition at $t_c$ after core bounce. {\bf a} and {\bf b.} neutrino luminosities $L_\nu$ {\bf (a)} and average energies $\langle E_\nu \rangle$ {\bf (b)}, of all neutrino and antineutrino flavors of electron ($e$), muon ($\mu$) and tau ($\tau$), sampled in the co-moving frame of reference at 1000~km. In {\bf (a)} we skip $\bar\nu_{\mu/\tau}$ since $L_{\nu_{\mu/\tau}}\simeq L_{\bar\nu_{\mu/\tau}}$. {\bf c.} neutrino signal analysis at the Super-Kamiokande detector for a fiducial event at a distance of 10~kpc, taking into account possible effects of neutrino flavour oscillations (pink region, see text for details). The blue crosses show the result in the normal neutrino mass hierarchy.

\newpage

{\Large \bf \hspace{-12mm} Figures} \\
\begin{figure*}[htp!]
\resizebox{1.0\textwidth}{!}{\includegraphics{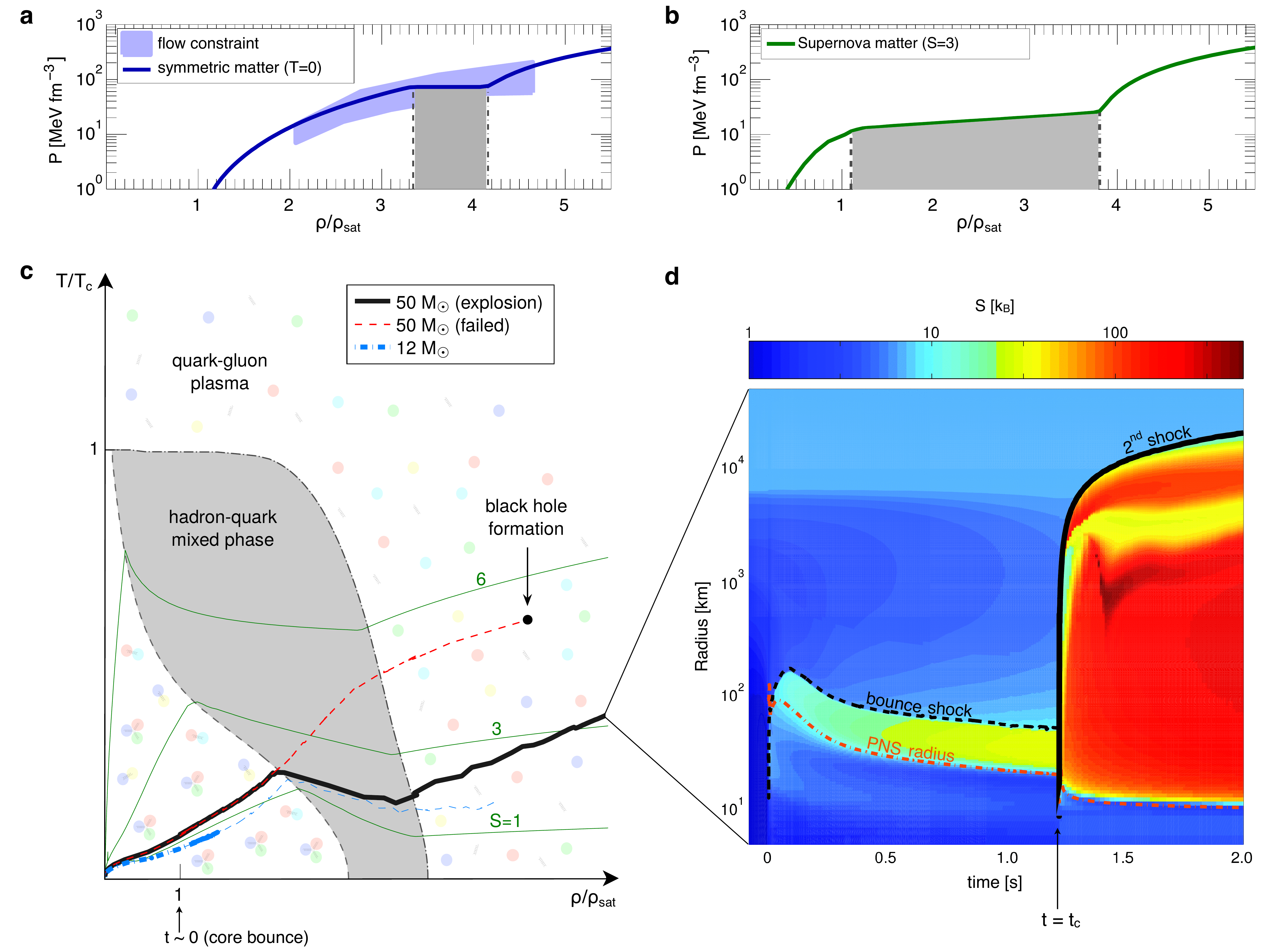}}
{\bf Figure~1.}~$\vert$ {\bf Hybrid EOS, phase diagram and supernova evolution.}
\label{fig1}
\end{figure*}
\\
\begin{figure*}
\resizebox{0.5\textwidth}{!}{\includegraphics{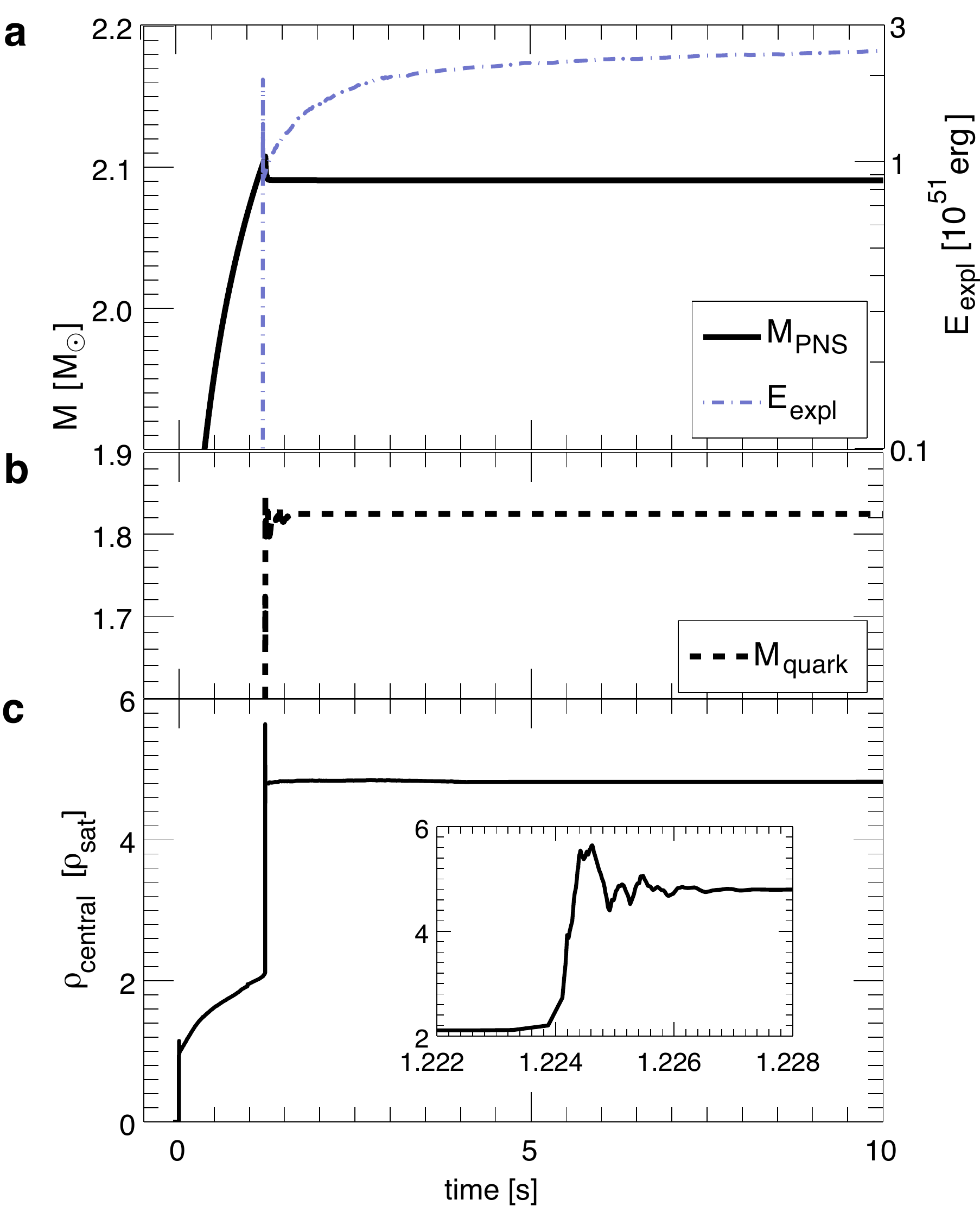}}
\\
{\bf Figure~2.}~$\vert$ {\bf Evolution of the supernova simulation}
\label{fig2}
\end{figure*}
\\ \\ \\
\begin{figure*}
\resizebox{0.5\textwidth}{!}{\includegraphics{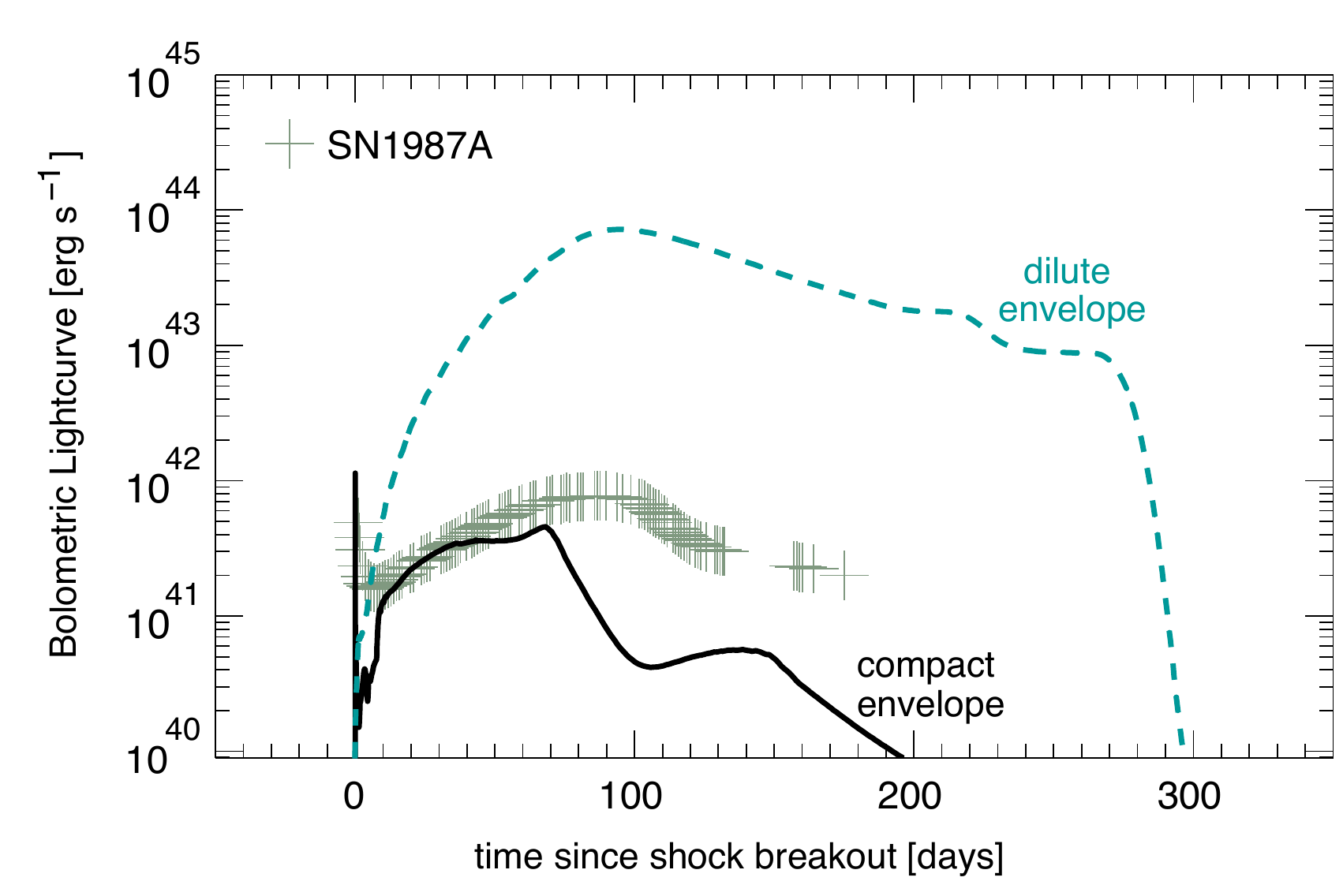}}
\\
{\bf Figure~3.}~$\vert$ {\bf Bolometric light curve}
\label{fig3}
\end{figure*}
\\ \\ \\
\begin{figure*}
\resizebox{0.5\textwidth}{!}{\includegraphics{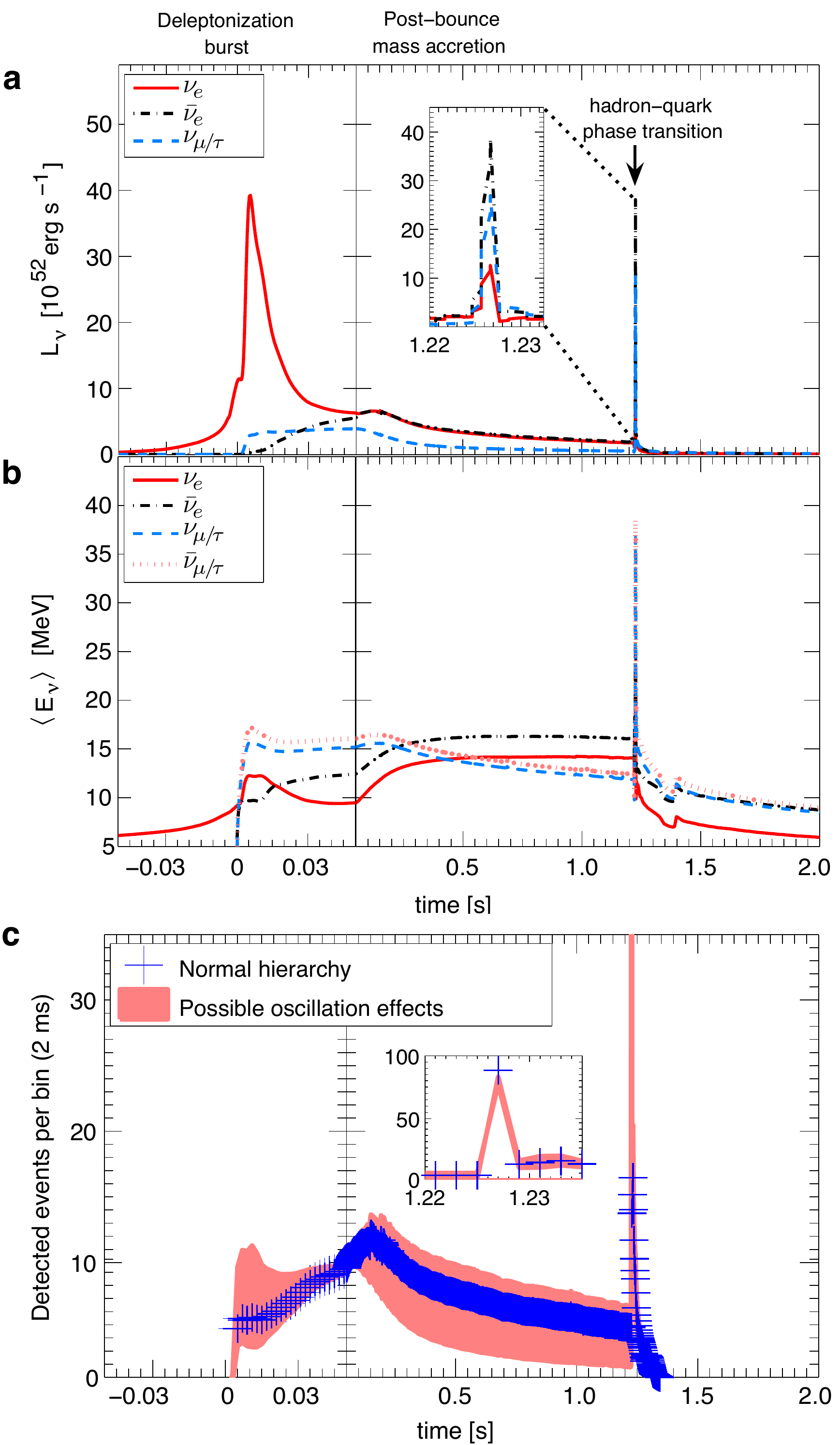}}
\\
{\bf Figure.~4.}~$\vert$ {\bf Neutrinos}
\label{fig4}
\end{figure*}

\newpage

{\Large \bf \hspace{-12mm} Supplementary Material} \\ \\
\hspace{-10mm}In the Supplementary Figure~1, we provide the sequences of neutron star configurations based on the newly developed SF EOS parametrisation (black lines), in comparison to the hadronic reference EOS (red dash-dotted lines)\cite{Typel:2009sy}. It contains the neutron star mass-radius relations in graph~(a) as well as the corresponding neutron star matter EOS, pressure and sound speed with respect to the energy density in graph~(b) -- only shown for the SF EOS. The tidal deformabilities are displayed in graphs~(c) and (d). The latter are compared with the data obtained by the LIGO/Virgo collaboration\cite{Abbott:2017} from the first observation of a binary neutron star merger GW170817. The employed low-mass prior neutron star masses, $m_1=1.36-1.60$~M$_\odot$ and $m_2=1.16-1.36$~M$_\odot$, corresponding to the tidal deformabilities $\Lambda_1$ and $\Lambda_2$ respectively, which are shown in graph~(c).
\begin{figure*}[h!]
\centering
\resizebox{1.0\textwidth}{!}{\includegraphics{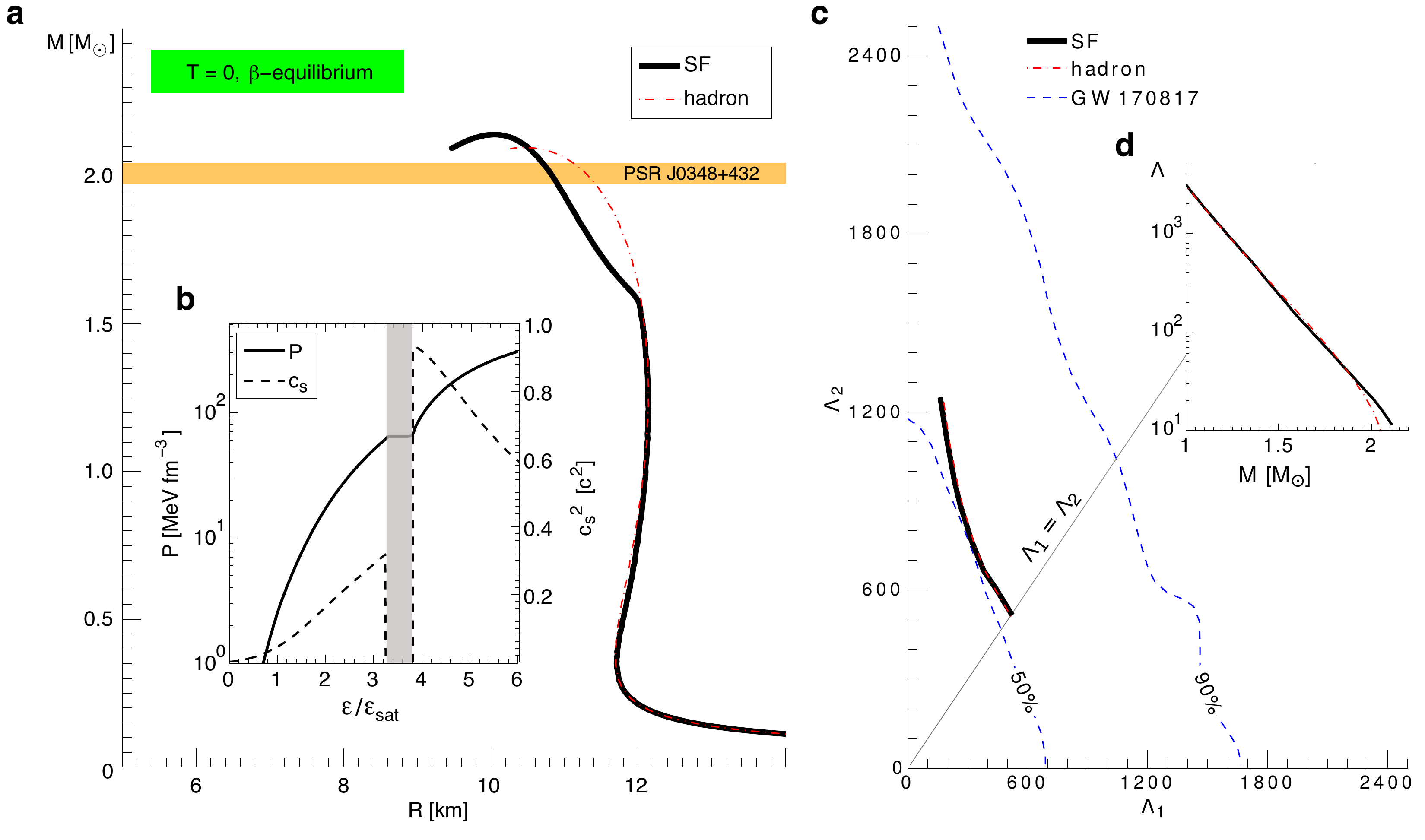}}
\\
{\bf Supplementary Figure 1.}~$\vert$~{{\bf Neutron star configurations} comparing the SF EOS (black lines) and the underlying hadronic EOS\cite{Typel:2009sy} (red dash-dotted line). {\bf a} mass-radius relations. The horizontal dark-yellow band marks the maximum neutron star mass constraint deduced from the pulsar observation PSR~J0348+432\cite{Antoniadis:2013}. {\bf b.} pressure $P$ (solid line) and square of the sound speed $c_s$ in units of the speed of light $c$ (dashed line) of the SF EOS only, with respect of the energy density $\varepsilon$ expressed in units of the value at saturation density, $\varepsilon_{\rm sat}\simeq 142$~MeV~fm$^{-3}$ in $\beta$-equilibrium at zero temperature. The grey-shaded region marks the hadron-quark mixed phase.  {\bf c.} $\Lambda_1$--$\Lambda_2$ tidal deformability relation as constrained from GW170817 (blue dashed lines show the 50\% and 90\% confidence levels) corresponding to the low-mass prior neutron star masses\cite{Abbott:2017}, and {\bf d.} tidal deformability dependence on the neutron star mass.}
\label{figSup1}
\end{figure*}
\\ \\
Note that the maximum hybrid star mass, $M_{\rm max}=2.17~$M$_\odot$, predicted by the SF EOS not only satisfies the lower limit of $2.01$~M$_\odot$ (see graph~(a) in the Supplementary Figure~1), but also fulfills the upper limit of $\sim2.2$~M$_\odot$, recently deduced from the multi-messenger observation of GW170817\cite{Margalit:2017dij,Shibata:2017xdx,Rezzolla:2017aly}.
\\
The behaviour of the speed of sound is largely unknown between the maximum (energy) density of about twice the saturation value -- up to which ab-initio nuclear matter calculations might at best be extrapolated, e.g., chiral effective field theories -- and the maximal densities attained in the core of neutron stars, which are still well below the asymptotics of the conformal limit case $c_s^2 = 1/3$. In a recent work\cite{Tews:2018iwm} different cases have been discussed that would fulfill the observational constraints for the compactness and the maximum mass of compact stars but that do not cover the case of a first-order phase transition which is provided by our hadron-quark hybrid star EOS. This work suggests that any hadron-quark hybrid EOS which satisfies both the current nuclear physics constraint at $\rho<2\times\rho_{\rm sat}$ and the lower bound on the maximum mass of $2.01$~M$_\odot$, would have a region where the sound speed exceeds the conformal limit. We note here that an EOS construction which switches from a well constrained nuclear EOS directly to an asymptotic EOS, that obeys the conformal limit with $c_s^2=1/3$, would fulfill the 2~M$_\odot$ constraint on the maximum mass, provided the transition occurs at low enough densities not exceeding $2\times\rho_{\rm sat}$\cite{Alford:2015dpa}. A detailed discussion of this statement and a variety of model ansaetze for $c_s^2$ can be found in the recent literature\cite{Tews:2018iwm}. In our SF EOS that features a first-order phase transition, which was not considered previously\cite{Tews:2018iwm}, the value of the sound speed right after the phase-transition exceeds this limit but gradually approaches it at large energy density. \\
\newpage
\begin{figure*}[h!]
\centering
\resizebox{1.0\textwidth}{!}{\includegraphics{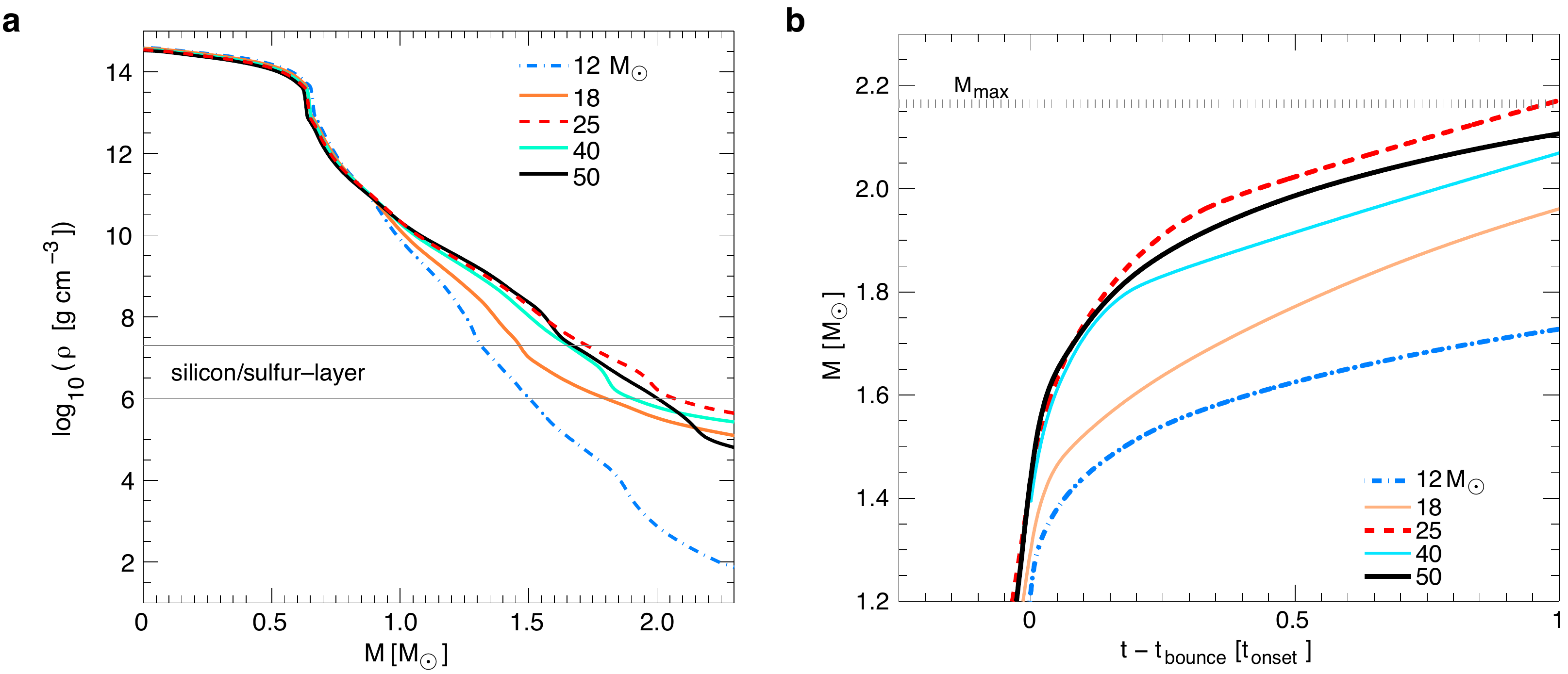}}
\\
{\bf Supplementary Figure 2.}~$\vert$~{{\bf Supernova simulation summary} of all stellar models considered, ZAMS masses of 12--40~M$_\odot$\cite{Woosley:2002zz} and 50~M$_\odot$\cite{Umeda:2007wk}, where the quark-hadron phase transition was taken into account based on the SF EOS. {\bf a} restmass density profile at the moment of core bounce, $t_{\rm bounce}$. The horizontal band approximately marks the region of the silicon/sulfur layer, which determines the mass accretion onto the central proto-neutron star during the post-bounce evolution. {\bf b} post-bounce evolution of the baryon mass enclosed inside the PNS, relative the onset times, $t_{\rm onset}$, when the hadron-quark phase boundary is reached (the corresponding conditions are given in Table~1 of the manuscript). The medium-modified SF EOS maximum mass is approximated by the horizontal band denoted as $M_{\rm max}$.}
\label{figSup2}
\end{figure*}


\begin{thebibliography}{}
\expandafter\ifx\csname url\endcsname\relax
  \def\url#1{\texttt{#1}}\fi
\expandafter\ifx\csname urlprefix\endcsname\relax\def\urlprefix{URL }\fi
\providecommand{\bibinfo}[2]{#2}
\providecommand{\eprint}[2][]{\url{#2}}

\end{thebibliography}


\begin{thebibliography}{60}
\expandafter\ifx\csname url\endcsname\relax
  \def\url#1{\texttt{#1}}\fi
\expandafter\ifx\csname urlprefix\endcsname\relax\def\urlprefix{URL }\fi
\providecommand{\bibinfo}[2]{#2}
\providecommand{\eprint}[2][]{\url{#2}}

\bibitem{Janka:2007}
\bibinfo{author}{{Janka}, H.-T.}, \bibinfo{author}{{Langanke}, K.}, \bibinfo{author}{{Marek}, A.}, \bibinfo{author}{{Mart{\'{\i}}nez-Pinedo}, G.} \& \bibinfo{author}{{M{\"u}ller}, B.}
\newblock \bibinfo{title}{{Theory of core-collapse supernovae}}.
\newblock \textit{ \bibinfo{journal}{\physrep}} \textbf{\bibinfo{volume}{442}}, \bibinfo{pages}{38-74} (\bibinfo{year}{2007}).

\bibitem{GalYam:2009}
\bibinfo{author}{{Gal-Yam}, A.} \& \bibinfo{author}{{Leonard}, D.~C.}
\newblock \bibinfo{title}{{A massive hypergiant star as the progenitor of the supernova SN 2005gl}}.
\newblock \textit{ \bibinfo{journal}{\nat}} \textbf{\bibinfo{volume}{458}}, \bibinfo{pages}{865-867} (\bibinfo{year}{2009}).

\bibitem{Foley:2011}
\bibinfo{author}{{Foley}, R.~J.} \textit{ et~al.}
\newblock \bibinfo{title}{{The Diversity of Massive Star Outbursts. I. Observations of SN2009ip, UGC 2773 OT2009-1, and Their Progenitors}}.
\newblock \textit{ \bibinfo{journal}{\apj}} \textbf{\bibinfo{volume}{732}}, \bibinfo{pages}{32} (\bibinfo{year}{2011}).

\bibitem{Zhang:2012}
\bibinfo{author}{{Zhang}, T.} \textit{ et~al.}
\newblock \bibinfo{title}{{Type IIn Supernova SN 2010jl: Optical Observations for over 500 Days after Explosion}}.
\newblock \textit{ \bibinfo{journal}{\aj}} \textbf{\bibinfo{volume}{144}}, \bibinfo{pages}{131} (\bibinfo{year}{2012}).

\bibitem{Mauerhan:2013}
\bibinfo{author}{{Mauerhan}, J.~C.} \textit{ et~al.}
\newblock \bibinfo{title}{{The unprecedented 2012 outburst of SN 2009ip: a luminous blue variable star becomes a true supernova}}.
\newblock \textit{ \bibinfo{journal}{\mnras}} \textbf{\bibinfo{volume}{430}}, \bibinfo{pages}{1801-1810} (\bibinfo{year}{2013}).

\bibitem{Nicholl:2013}
\bibinfo{author}{{Nicholl}, M.} \textit{ et~al.}
\newblock \bibinfo{title}{{Slowly fading super-luminous supernovae that are not pair-instability explosions}}.
\newblock \textit{ \bibinfo{journal}{\nat}} \textbf{\bibinfo{volume}{502}}, \bibinfo{pages}{346-349} (\bibinfo{year}{2013}).

\bibitem{Terreran:2017}
\bibinfo{author}{{Terreran}, M.} \textit{ et~al.}
\newblock \bibinfo{title}{{Hydrogen-rich supernovae beyond the neutrino-driven core-collapse paradigm}}.
\newblock \textit{ \bibinfo{journal}{\nat}} \textbf{\bibinfo{volume}{1}}, \bibinfo{pages}{713-720} (\bibinfo{year}{2017}).

\bibitem{Sumiyoshi:2006id}
\bibinfo{author}{Sumiyoshi, K.}, \bibinfo{author}{Yamada, S.},
  \bibinfo{author}{Suzuki, H.} \& \bibinfo{author}{Chiba, S.}
\newblock \bibinfo{title}{{Neutrino signals from the formation of black hole: A probe of equation of state of dense matter}}.
\newblock \textit{ \bibinfo{journal}{Phys.Rev.Lett.}} \textbf{\bibinfo{volume}{97}}, \bibinfo{pages}{091101} (\bibinfo{year}{2006}).

\bibitem{Fischer:2009}
\bibinfo{author}{{Fischer}, T.}, \bibinfo{author}{{Whitehouse}, S.~C.}, \bibinfo{author}{{Mezzacappa}, A.}, \bibinfo{author}{{Thielemann}, F.-K.} \& \bibinfo{author}{{Liebend{\"o}rfer}, M.}
\newblock \bibinfo{title}{{The neutrino signal from protoneutron star accretion and black hole formation}}.
\newblock \textit{ \bibinfo{journal}{\aap}} \textbf{\bibinfo{volume}{499}}, \bibinfo{pages}{1-15} (\bibinfo{year}{2009}).

\bibitem{OConnor:2011}
\bibinfo{author}{{O'Connor}, E.} \& \bibinfo{author}{{Ott}, C.~D.}
\newblock \bibinfo{title}{{Black Hole Formation in Failing Core-Collapse Supernovae}}.
\newblock \textit{ \bibinfo{journal}{\apj}} \textbf{\bibinfo{volume}{730}}, \bibinfo{pages}{70} (\bibinfo{year}{2011}).

\bibitem{Mueller:2017}
\bibinfo{author}{{Chan}, C.}, \bibinfo{author}{{M{\"u}ller}, B.}, \bibinfo{author}{{Heger}, A.}, \bibinfo{author}{{Pakmor}, R.} \& \bibinfo{author}{{Springel}, V.}
\newblock \bibinfo{title}{{Black hole formation and fallback during the supernova explosion of a 40~$M_\odot$ star}}.
\newblock \textit{ \bibinfo{journal}{\apj}} \textbf{\bibinfo{volume}{852}}, \bibinfo{pages}{L19} (\bibinfo{year}{2017}).

\bibitem{Woosley:2002zz}
\bibinfo{author}{Woosley, S.}, \bibinfo{author}{Heger, A.} \& \bibinfo{author}{Weaver, T.}
\newblock \bibinfo{title}{{The evolution and explosion of massive stars}}.
\newblock \textit{ \bibinfo{journal}{Rev.Mod.Phys.}} \textbf{\bibinfo{volume}{74}}, \bibinfo{pages}{1015-1071} (\bibinfo{year}{2002}).

\bibitem{Umeda:2007wk}
\bibinfo{author}{Umeda, H.} \& \bibinfo{author}{Nomoto, K.}
\newblock \bibinfo{title}{{How much Ni-56 can be produced in Core-Collapse Supernovae? Evolution and Explosions of 30--100 M$_\odot$ Stars}}.
\newblock \textit{ \bibinfo{journal}{Astrophys.J.}} \textbf{\bibinfo{volume}{673}}, \bibinfo{pages}{1014} (\bibinfo{year}{2008}).

\bibitem{Antoniadis:2013}
\bibinfo{author}{{Antoniadis}, J.} \textit{ et~al.}
\newblock \bibinfo{title}{{A Massive Pulsar in a Compact Relativistic Binary}}.
\newblock \textit{ \bibinfo{journal}{Science}} \textbf{\bibinfo{volume}{340}}, \bibinfo{pages}{448} (\bibinfo{year}{2013}).

\bibitem{Fonseca:2016}
\bibinfo{author}{{Fonseca}, E.} \textit{ et~al.}
\newblock \bibinfo{title}{{The NANOGrav Nine-year Data Set: Mass and Geometric Measurements of Binary Millisecond Pulsars}}.
\newblock \textit{ \bibinfo{journal}{\apj}} \textbf{\bibinfo{volume}{832}}, \bibinfo{pages}{167} (\bibinfo{year}{2016}).

\bibitem{Liebendoerfer:2004}
\bibinfo{author}{Liebend\"orfer, M.} \textit{ et~al.}
\newblock \bibinfo{title}{{A Finite difference representation of neutrino radiation hydrodynamics for spherically symmetric general relativistic supernova simulations}}.
\newblock \textit{ \bibinfo{journal}{Astrophys.J.Suppl.}} \textbf{\bibinfo{volume}{150}}, \bibinfo{pages}{263} (\bibinfo{year}{2004}).

\bibitem{Typel:2009sy}
\bibinfo{author}{Typel, S.}, \bibinfo{author}{R{\"o}pke, G.}, \bibinfo{author}{Kl{\"a}hn, T.}, \bibinfo{author}{Blaschke, D.} \& \bibinfo{author}{Wolter, H.}
\newblock \bibinfo{title}{{Composition and thermodynamics of nuclear matter with light clusters}}.
\newblock \textit{ \bibinfo{journal}{Phys.Rev.}} \textbf{\bibinfo{volume}{C81}}, \bibinfo{pages}{015803} (\bibinfo{year}{2010}).

\bibitem{Danielewicz:2002}
\bibinfo{author}{{Danielewicz}, P.}, \bibinfo{author}{{Lacey}, R.} \& \bibinfo{author}{{Lynch}, W.~G.}
\newblock \bibinfo{title}{{Determination of the Equation of State of Dense Matter}}.
\newblock \textit{ \bibinfo{journal}{Science}} \textbf{\bibinfo{volume}{298}}, \bibinfo{pages}{1592} (\bibinfo{year}{2002}).

\bibitem{Lattimer:2013}
\bibinfo{author}{{Lattimer}, J.~M.} \& \bibinfo{author}{{Lim}, Y.}
\newblock \bibinfo{title}{{Constraining the Symmetry Parameters of the Nuclear Interaction}}.
\newblock \textit{ \bibinfo{journal}{\apj}} \textbf{\bibinfo{volume}{771}}, \bibinfo{pages}{51} (\bibinfo{year}{2013}).

\bibitem{Krueger:2013}
\bibinfo{author}{{Kr{\"u}ger}, T.}, \bibinfo{author}{{Tews}, I.}, \bibinfo{author}{{Hebeler}, K.} \& \bibinfo{author}{{Schwenk}, A.}
\newblock \bibinfo{title}{{Neutron matter from chiral effective field theory interactions}}.
\newblock \textit{ \bibinfo{journal}{Phys.Rev.}} \textbf{\bibinfo{volume}{C88}}, \bibinfo{pages}{025802} (\bibinfo{year}{2013}).

\bibitem{Laermann:2012PRL}
\bibinfo{author}{{Bazavov}, A.} \textit{ et~al.} [HotQCD collaboration]
\newblock \bibinfo{title}{{Equation of state in (2+1)-flavor QCD}}.
\newblock \textit{ \bibinfo{journal}{Phys.Rev.}} \textbf{\bibinfo{volume}{D90}}, \bibinfo{pages}{094503} (\bibinfo{year}{2014}).

\bibitem{Katz:2014PhLB}
\bibinfo{author}{{Bors{\'a}nyi}, S.} \textit{ et~al.}
\newblock \bibinfo{title}{{Full result for the QCD equation of state with 2+1 flavors}}.
\newblock \textit{ \bibinfo{journal}{Phys.Lett.}} \textbf{\bibinfo{volume}{B730}}, \bibinfo{pages}{99} (\bibinfo{year}{2014}).

\bibitem{Bazavov:2011nk}
\bibinfo{author}{{Bazavov}, A.} \textit{ et~al.}
\newblock \bibinfo{title}{{The chiral and deconfinement aspects of the QCD transition}}.
\newblock \textit{ \bibinfo{journal}{Phys.Rev.}} \textbf{\bibinfo{volume}{D85}}, \bibinfo{pages}{054503} (\bibinfo{year}{2012}).

\bibitem{Kurkela:2014vha}
\bibinfo{author}{Kurkela, A.}, \bibinfo{author}{Fraga, E.~S.}, \bibinfo{author}{Schaffner-Bielich, J.} \& \bibinfo{author}{Vuorinen, A.}
\newblock \bibinfo{title}{{Constraining neutron star matter with Quantum  Chromodynamics}}.
\newblock \textit{ \bibinfo{journal}{\apj}} \textbf{\bibinfo{volume}{789}}, \bibinfo{pages}{127} (\bibinfo{year}{2014}).

\bibitem{Farhi:1984qu}
\bibinfo{author}{Farhi, E.} \& \bibinfo{author}{Jaffe, R.}
\newblock \bibinfo{title}{{Strange Matter}}. 
\newblock \textit{ \bibinfo{journal}{Phys.Rev.}} \textbf{\bibinfo{volume}{D30}}, \bibinfo{pages}{2379} (\bibinfo{year}{1984}).

\bibitem{Nambu:1961tp}
\bibinfo{author}{Nambu, Y.} \& \bibinfo{author}{Jona-Lasinio, G.}
\newblock \bibinfo{title}{{Dynamical Model of Elementary Particles Based on an Analogy with Superconductivity. 1.}}
\newblock \textit{ \bibinfo{journal}{Phys.Rev.}} \textbf{\bibinfo{volume}{122}}, \bibinfo{pages}{345} (\bibinfo{year}{1961}).

\bibitem{Takahara:1988yd}
\bibinfo{author}{Takahara, M.}  \& \bibinfo{author}{Sato, K.}
\newblock \bibinfo{title}{{Phase transition in the newly born neutron star and neutrino emission from SN1987A}}.
\newblock \textit{ \bibinfo{journal}{Prog.Theor.Phys.}} \textbf{\bibinfo{volume}{80}}, \bibinfo{pages}{861-867} (\bibinfo{year}{1988}).

\bibitem{Gentile:1993ma}
\bibinfo{author}{Gentile, N A.}, \bibinfo{author}{Aufderheide, M. B.}, \bibinfo{author}{Mathews, G. J.},  \bibinfo{author}{Swesty, F. D}, \&  \bibinfo{author}{Fuller, G. M.}, 
\newblock \bibinfo{title}{{The QCD phase transition and supernova core collapse}}.
\newblock \textit{ \bibinfo{journal}{\apj}} \textbf{\bibinfo{volume}{414}}, \bibinfo{pages}{701} (\bibinfo{year}{1993}).

\bibitem{Sagert:2008ka}
\bibinfo{author}{Sagert, I.} \textit{ et~al.}
\newblock \bibinfo{title}{{Signals of the QCD phase transition in core-collapse supernovae}}.
\newblock \textit{ \bibinfo{journal}{Phys.Rev.Lett.}} \textbf{\bibinfo{volume}{102}}, \bibinfo{pages}{081101} (\bibinfo{year}{2009}).

\bibitem{Nakazato:2008su}
\bibinfo{author}{{Nakazato}, K.}, \bibinfo{author}{{Sumiyoshi}, K.} \& \bibinfo{author}{{Yamada}, S.},
\newblock \bibinfo{title}{{Astrophysical Implications of Equation of State for Hadron-Quark Mixed Phase: Compact Stars and Stellar Collapses}}.
\newblock \textit{ \bibinfo{journal}{Phys.Rev.}} \textbf{\bibinfo{volume}{D77}}, \bibinfo{pages}{103006} (\bibinfo{year}{2008}).

\bibitem{Blaschke:2017}
\bibinfo{author}{{Kaltenborn}, M.~A.~R.}, \bibinfo{author}{{Bastian}, N.-U.~F.} \& \bibinfo{author}{{Blaschke}, D.~B.}
\newblock \bibinfo{title}{{Quark-nuclear hybrid star equation of state with excluded volume effects}}.
\newblock \textit{ \bibinfo{journal}{Phys.Rev.}} \textbf{\bibinfo{volume}{D96}}, \bibinfo{pages}{056024} (\bibinfo{year}{2017}).

\bibitem{Abbott:2017}
\bibinfo{author}{{Abbott}, B.~P.} \textit{ et~al.} [LIGO scientific and Virgo collaborations]
\newblock \bibinfo{title}{{GW170817: Observation of Gravitational Waves from a Binary Neutron Star Inspiral}}.
\newblock \textit{ \bibinfo{journal}{Phys.Rev.Lett.}} \textbf{\bibinfo{volume}{119}}, \bibinfo{pages}{161101} (\bibinfo{year}{2017}).

\bibitem{Lattimer:2018}
\bibinfo{author}{{De}, S.}, \bibinfo{author}{{Finstad}, D.},  \bibinfo{author}{{Lattimer}, J.~M.}, \bibinfo{author}{{Brown}, D.~A.}, \bibinfo{author}{{Berger}, E.} \& \bibinfo{author}{{Biwer}, C.~M.}
\newblock \bibinfo{title}{{Constraining the nuclear equation of state with GW170817}}.
\newblock \textit{ \bibinfo{journal}{\prl}} in press. (\bibinfo{year}{2018}).

\bibitem{Horowitz:1985}
\bibinfo{author}{{Horowitz}, C.~J.}, \bibinfo{author}{{Moniz}, E.~J.} \& \bibinfo{author}{{Negele}, J.~W.}
\newblock \bibinfo{title}{{Hadron structure in a simple model of quark/nuclear matter}}.
\newblock \textit{ \bibinfo{journal}{Phys.Rev.}} \textbf{\bibinfo{volume}{D31}}, \bibinfo{pages}{1689} (\bibinfo{year}{1985}).

\bibitem{Blaschke:1986}
\bibinfo{author}{{R{\"o}pke}, G.}, \bibinfo{author}{{Blaschke}, D.} \& \bibinfo{author}{{Schulz}, H.}
\newblock \bibinfo{title}{{Pauli quenching effects in a simple string model of quark/nuclear matter}}.
\newblock \textit{ \bibinfo{journal}{Phys.Rev.}} \textbf{\bibinfo{volume}{D34}},\bibinfo{pages}{3499} (\bibinfo{year}{1986}).

\bibitem{Klaehn:2015}
\bibinfo{author}{{Kl{\"a}hn}, T.} \& \bibinfo{author}{{Fischer}, T.}
\newblock \bibinfo{title}{{Vector Interaction Enhanced Bag Model for Astrophysical Applications}}.
\newblock \textit{ \bibinfo{journal}{\apj}} \textbf{\bibinfo{volume}{810}}, \bibinfo{pages}{134} (\bibinfo{year}{2015}).

\bibitem{Benic:2014jia}
\bibinfo{author}{Beni{\'c}, S.}, \bibinfo{author}{Blaschke, D.}, \bibinfo{author}{Alvarez-Castillo, D.~E.}, \bibinfo{author}{Fischer, T.} \& \bibinfo{author}{Typel, S.}
\newblock \bibinfo{title}{{A new quark-hadron hybrid equation of state for astrophysics - I. High-mass twin compact stars}}.
\newblock \textit{ \bibinfo{journal}{\aap}} \textbf{\bibinfo{volume}{577}}, \bibinfo{pages}{A40} (\bibinfo{year}{2015}).

\bibitem{Bethe:1985ux}
\bibinfo{author}{Bethe, H.~A.} \& \bibinfo{author}{Wilson, R., James}.
\newblock \bibinfo{title}{{Revival of a stalled supernova shock by neutrino heating}}.
\newblock \textit{ \bibinfo{journal}{\apj}} \textbf{\bibinfo{volume}{295}}, \bibinfo{pages}{14} (\bibinfo{year}{1985}).

\bibitem{LeBlanc:1970kg}
\bibinfo{author}{LeBlanc, J.~M.} \& \bibinfo{author}{Wilson, J.~R.}
\newblock \bibinfo{title}{{A Numerical Example of the Collapse of a Rotating Magnetized Star}}.
\newblock \textit{ \bibinfo{journal}{\apj}} \textbf{\bibinfo{volume}{161}}, \bibinfo{pages}{541} (\bibinfo{year}{1970}).

\bibitem{Bisnovatyi-Kogan:1976}
\bibinfo{author}{{Bisnovatyi-Kogan}, G.~S.}, \bibinfo{author}{{Popov}, I.~P.} \& \bibinfo{author}{{Samokhin}, A.~A.}
\newblock \bibinfo{title}{{The magnetohydrodynamic rotational model of supernova explosion}}.
\newblock \textit{ \bibinfo{journal}{\apss}} \textbf{\bibinfo{volume}{41}}, \bibinfo{pages}{287-320} (\bibinfo{year}{1976}).

\bibitem{Bruenn:2013}
\bibinfo{author}{{Bruenn}, S. W.}  \textit{ et~al.}
\newblock \bibinfo{title}{{Axisymmetric Ab Initio Core-collapse Supernova Simulations of 12--25 M $_\odot$ Stars}}.
\newblock \textit{ \bibinfo{journal}{\apjl}} \textbf{\bibinfo{volume}{767}}, \bibinfo{pages}{L6} (\bibinfo{year}{2013}).

\bibitem{Mueller:2012a}
\bibinfo{author}{{M{\"u}ller}, B.}, \bibinfo{author}{{Janka}, H.-Th.} \&  \bibinfo{author}{{Marek}, A.}
\newblock \bibinfo{title}{{A New Multi-dimensional General Relativistic Neutrino Hydrodynamics Code for Core-collapse Supernovae. II. Relativistic Explosion Models of Core-collapse Supernovae}}.
\newblock \textit{ \bibinfo{journal}{\apj}} \textbf{\bibinfo{volume}{756}}, \bibinfo{pages}{84} (\bibinfo{year}{2012}).

\bibitem{Suwa:2013}
\bibinfo{author}{{Suwa}, Y.}, \bibinfo{author}{{Takiwaki}, T.}, \bibinfo{author}{{Kotake}, K.}, \bibinfo{author}{{Fischer}, T.}, \bibinfo{author}{{Liebend{\"o}rfer}, M.} \&  \bibinfo{author}{{Sato}, K.}
\newblock \bibinfo{title}{{On the Importance of the Equation of State for the Neutrino-driven Supernova Explosion Mechanism}}.
\newblock \textit{ \bibinfo{journal}{\apj}} \textbf{\bibinfo{volume}{764}}, \bibinfo{pages}{99} (\bibinfo{year}{2013}).

\bibitem{Melson:2015}
\bibinfo{author}{{Melson}, T.}, \bibinfo{author}{{Janka}, H.-Th.} \&  \bibinfo{author}{{Marek}, A.}
\newblock \bibinfo{title}{{Neutrino-driven Supernova of a Low-mass Iron-core Progenitor Boosted by Three-dimensional Turbulent Convection}}.
\newblock \textit{ \bibinfo{journal}{\apjl}} \textbf{\bibinfo{volume}{801}}, \bibinfo{pages}{L24} (\bibinfo{year}{2015}).

\bibitem{Wu:2016}
\bibinfo{author}{{Wu}, M.-R.}, \bibinfo{author}{{Fern{\'a}ndez}, R.}, \bibinfo{author}{{Mart{\'{\i}}nez-Pinedo}, G.} \& \bibinfo{author}{{Metzger}, B.~D.}
\newblock \bibinfo{title}{{Production of the entire range of r-process nuclides by black hole accretion disc outflows from neutron star mergers}}.
\newblock \textit{ \bibinfo{journal}{\mnras}} \textbf{\bibinfo{volume}{463}}, \bibinfo{pages}{2323-2334} (\bibinfo{year}{2016}).

\bibitem{Smith:2007}
\bibinfo{author}{{Smith}, N.}, \bibinfo{author}{{Li}, W.}, \bibinfo{author}{{Foley}, R.~J.}  \textit{ et~al.} 
\newblock \bibinfo{title}{{SN 2006gy: Discovery of the Most Luminous Supernova Ever Recorded, Powered by the Death of an Extremely Massive Star like {$\eta$} Carinae}}.
\newblock \textit{ \bibinfo{journal}{\apj}} \textbf{\bibinfo{volume}{666}}, \bibinfo{pages}{1116} (\bibinfo{year}{2007}).

\bibitem{Blinnikov:2012}
\bibinfo{author}{{Moriya}, T. J.}, \bibinfo{author}{{Blinnikov}, S.~I.}, \bibinfo{author}{{Tominaga}, N.}, \bibinfo{author}{{Yoshida}, N.}, \bibinfo{author}{{Tanaka}, M.}, \bibinfo{author}{{Maeda}, K.} \& \bibinfo{author}{{Nomoto}, K.}
\newblock \bibinfo{title}{{Light-curve modelling of superluminous supernova 2006gy: collision between supernova ejecta and a dense circumstellar medium}}.
\newblock \textit{ \bibinfo{journal}{\mnras}} \textbf{\bibinfo{volume}{428}}, \bibinfo{pages}{1020-1035} (\bibinfo{year}{2013}).

\bibitem{Mirizzi:2016}
\bibinfo{author}{{Mirizzi}, A.} \textit{ et~al.}
\newblock \bibinfo{title}{{Supernova neutrinos: production, oscillations and detection}}.
\newblock \textit{ \bibinfo{journal}{La Rivista del Nuovo Cimento}} \textbf{\bibinfo{volume}{39}}, \bibinfo{pages}{1-112} (\bibinfo{year}{2016}).

\bibitem{Dasgupta:2009yj}
\bibinfo{author}{Dasgupta, B.} \textit{ et~al.}
\newblock \bibinfo{title}{{Detecting the QCD phase transition in the next Galactic supernova neutrino burst}}.
\newblock \textit{ \bibinfo{journal}{Phys.Rev.}} \textbf{\bibinfo{volume}{D81}}, \bibinfo{pages}{103005} (\bibinfo{year}{2010}).

\bibitem{Tauris:2011}
\bibinfo{author}{{Tauris}, T.}, \bibinfo{author}{{Langer}, N.},  \& \bibinfo{author}{{Kramer}, M.}
\newblock \bibinfo{title}{{Formation of millisecond pulsars with CO white dwarf companions - I. PSR J1614-2230: evidence for a neutron star born massive}}.
\newblock \textit{ \bibinfo{journal}{\mnras}} \textbf{\bibinfo{volume}{416}}, \bibinfo{pages}{2130-2142} (\bibinfo{year}{2011}).

\bibitem{Tauris:2012}
\bibinfo{author}{{Tauris}, T.}, \bibinfo{author}{{Langer}, N.},  \& \bibinfo{author}{{Kramer}, M.}
\newblock \bibinfo{title}{{Formation of millisecond pulsars with CO white dwarf companions - II. Accretion, spin-up, true ages and comparison to MSPs with He white dwarf companions}}.
\newblock \textit{ \bibinfo{journal}{\mnras}} \textbf{\bibinfo{volume}{425}}, \bibinfo{pages}{1601-1627} (\bibinfo{year}{2012}).

\bibitem{Sukhbold:2016}
\bibinfo{author}{{Sukhbold}, T.}, \bibinfo{author}{{Ertl}, T.}, \bibinfo{author}{{Woosley}, S.~E.},  \bibinfo{author}{{Brown}, J.~M.}  \& \bibinfo{author}{{Janka}, H.-Th.}
\newblock \bibinfo{title}{{Core-collapse Supernovae from 9 to 120 Solar Masses Based on Neutrino-powered Explosions}}.
\newblock \textit{ \bibinfo{journal}{\apj}} \textbf{\bibinfo{volume}{821}}, \bibinfo{pages}{45} (\bibinfo{year}{2016}).

\bibitem{Paczyski:1971}
\bibinfo{author}{{Paczy{\'n}ski}, B.}
\newblock \bibinfo{title}{{Evolutionary Processes in Close Binary Systems}}.
\newblock \textit{ \bibinfo{journal}{Ann.Rev.Astron.Astrophys.}} \textbf{\bibinfo{volume}{9}}, \bibinfo{pages}{183} (\bibinfo{year}{1971}).

\bibitem{Ozel:2012}
\bibinfo{author}{{{\"O}zel}, F.}, \bibinfo{author}{{Psaltis}, D.}, \bibinfo{author}{{Narayan}, R.} \& \bibinfo{author}{{Santos Villarreal}, A.}
\newblock \bibinfo{title}{{On the Mass Distribution and Birth Masses of Neutron Stars}}.
\newblock \textit{ \bibinfo{journal}{\apj}} \textbf{\bibinfo{volume}{757}}, \bibinfo{pages}{55} (\bibinfo{year}{2012}).

\bibitem{Antoniadis:2016}
\bibinfo{author}{{Antoniadis}, J.}, \bibinfo{author}{{Tauris}, M.~T.}, \bibinfo{author}{{{\"O}zel}, F.}, \bibinfo{author}{{Barr}, E.}, \bibinfo{author}{{Champion}, D.~J.}, \& \bibinfo{author}{{Freire}, P.~C.~C.}
\newblock \bibinfo{title}{{The millisecond pulsar mass distribution: Evidence for bimodality and constraints on the maximum neutron star mass}}.
\newblock \textit{ \bibinfo{journal}{ArXiv e-prints}}, astro-ph.HE/1605.01665 (\bibinfo{year}{2016}).

\bibitem{Mezzacappa:1993gn}
\bibinfo{author}{Mezzacappa, A.} \& \bibinfo{author}{Bruenn, S.}
\newblock \bibinfo{title}{{A numerical method for solving the neutrino Boltzmann equation coupled to spherically symmetric stellar core collapse}}.
\newblock \textit{ \bibinfo{journal}{\apj}} \textbf{\bibinfo{volume}{405}}, \bibinfo{pages}{669} (\bibinfo{year}{1993}).

\bibitem{Liebendoerfer:2002}
\bibinfo{author}{Liebendoerfer, M.}, \bibinfo{author}{Rosswog, S.} \& \bibinfo{author}{Thielemann, F.-K.}
\newblock \bibinfo{title}{{An Adaptive grid, implicit code for spherically symmetric, general relativistic hydrodynamics in comoving coordinates}}.
\newblock \textit{ \bibinfo{journal}{\apjs}} \textbf{\bibinfo{volume}{141}}, \bibinfo{pages}{229} (\bibinfo{year}{2002}).

\bibitem{Wu:2015}
\bibinfo{author}{{Wu}, M.-R.}, \bibinfo{author}{{Qian}, Y.-Z.}, \bibinfo{author}{{Mart{\'{\i}}nez-Pinedo}, G.}, \bibinfo{author}{{Fischer}, T.} \& \bibinfo{author}{{Huther}, L.}
\newblock \bibinfo{title}{{Effects of neutrino oscillations on nucleosynthesis and neutrino signals for an 18 M$_{\odot}$ supernova model}}.
\newblock \textit{ \bibinfo{journal}{Phys.Rev.}} \textbf{\bibinfo{volume}{D91}}, \bibinfo{pages}{065016} (\bibinfo{year}{2015}).

\bibitem{Scholberg:2012}
\bibinfo{author}{{Scholberg}, K.}
\newblock \bibinfo{title}{{Supernova Neutrino Detection}}.
\newblock \textit{ \bibinfo{journal}{Annual Review of Nuclear and Particle Science}} \textbf{\bibinfo{volume}{62}}, \bibinfo{pages}{81-103} (\bibinfo{year}{2012}).

\bibitem{PDG}
\bibinfo{author}{{Patrignani}, C.} \textit{ et~al.} (Particle Data Group)
\newblock \textit{ \bibinfo{journal}{Chin.Phys.}} \textbf{\bibinfo{volume}{C40}}, \bibinfo{pages}{100001} (\bibinfo{year}{2016}).

\bibitem{Blinnikov:2006}
\bibinfo{author}{{Blinnikov}, S. I.} \textit{ et~al.} 
%\bibinfo{author}{{R{\"o}pke}, F. K.}, \bibinfo{author}{{Sorokina}, E. I.},  \bibinfo{author}{{Gieseler}, M.},  \bibinfo{author}{{Reinecke}, M.},  \bibinfo{author}{{Travaglio}, C.}, \& \bibinfo{author}{{Hillebrandt}, W.}
\newblock \bibinfo{title}{{Theoretical light curves for deflagration models of type Ia supernova}}.
\newblock \textit{ \bibinfo{journal}{\aap}} \textbf{\bibinfo{volume}{453}}, \bibinfo{pages}{229-240} (\bibinfo{year}{2006}).

\bibitem{Sorokina:2016}
\bibinfo{author}{{Sorokina}, E.~I.}, \bibinfo{author}{{Blinnikov}, S.~I.}, \bibinfo{author}{{Nomoto}, K.}, \bibinfo{author}{{Quimby}, R.} \& \bibinfo{author}{{Tolstov}, A.}
\newblock \bibinfo{title}{{Type I Superluminous Supernovae as Explosions inside non-hydrogen Circumstellar Envelopes}}.
\newblock \textit{ \bibinfo{journal}{\apj}} \textbf{\bibinfo{volume}{829}}, \bibinfo{pages}{17} (\bibinfo{year}{2016})

\end{thebibliography}

\begin{thebibliography}{9}
\expandafter\ifx\csname url\endcsname\relax
  \def\url#1{\texttt{#1}}\fi
\expandafter\ifx\csname urlprefix\endcsname\relax\def\urlprefix{URL }\fi
\providecommand{\bibinfo}[2]{#2}
\providecommand{\eprint}[2][]{\url{#2}}

\bibitem{Typel:2009sy}
 \bibinfo{author}{Typel, S.},  \bibinfo{author}{R{\"o}pke, G.}  \bibinfo{author}{Kl{\"a}hn, T.}, \bibinfo{author}{Blaschke, D.} \& \bibinfo{author}{Wolter, H.}
\newblock \bibinfo{title}{{Composition and thermodynamics of nuclear matter with light clusters}}.
\newblock \textit{\bibinfo{journal}{Phys.Rev.}} \textbf{\bibinfo{volume}{C81}}, \bibinfo{pages}{015803} (\bibinfo{year}{2010}).

\bibitem{Abbott:2017}
\bibinfo{author}{{Abbott}, B.~P.} \textit{ et~al.} [LIGO scientific and Virgo collaborations]
\newblock \bibinfo{title}{{GW170817: Observation of Gravitational Waves from a Binary Neutron Star Inspiral}}.
\newblock \textit{ \bibinfo{journal}{Phys.Rev.Lett.}} \textbf{\bibinfo{volume}{119}}, \bibinfo{pages}{161101} (\bibinfo{year}{2017}).

\bibitem{Antoniadis:2013}
\bibinfo{author}{{Antoniadis}, J.} \textit{ et~al.}
\newblock \bibinfo{title}{{A Massive Pulsar in a Compact Relativistic Binary}}.
\newblock \textit{ \bibinfo{journal}{Science}} \textbf{\bibinfo{volume}{340}},
  \bibinfo{pages}{448} (\bibinfo{year}{2013}).

\bibitem{Margalit:2017dij}
\bibinfo{author}{Margalit, B.} \& \bibinfo{author}{Metzger, B.~D.}
\newblock \bibinfo{title}{{Constraining the Maximum Mass of Neutron Stars From
  Multi-Messenger Observations of GW170817}}.
\newblock \textit{ \bibinfo{journal}{\apj}} \textbf{\bibinfo{volume}{850}}, \bibinfo{pages}{L19} (\bibinfo{year}{2017}).

\bibitem{Shibata:2017xdx}
\bibinfo{author}{Shibata, M.} \textit{ et~al.}
\newblock \bibinfo{title}{{Modeling GW170817 based on numerical relativity and
  its implications}}.
\newblock \textit{ \bibinfo{journal}{Phys. Rev.}} \textbf{\bibinfo{volume}{D96}}, \bibinfo{pages}{123012} (\bibinfo{year}{2017}).

\bibitem{Rezzolla:2017aly}
\bibinfo{author}{Rezzolla, L.}, \bibinfo{author}{Most, E.~R.} \& \bibinfo{author}{Weih, L.~R.}
\newblock \bibinfo{title}{{Using gravitational-wave observations and
  quasi-universal relations to constrain the maximum mass of neutron stars}}.
\newblock \textit{ \bibinfo{journal}{\apj}} \textbf{\bibinfo{volume}{852}}, \bibinfo{pages}{L25} (\bibinfo{year}{2018}).

\bibitem{Tews:2018iwm}
\bibinfo{author}{{Tews}, I.}, \bibinfo{author}{{Carlson}, J.}, \bibinfo{author}{{Gandolfi}, S.} \& \bibinfo{author}{{Reddy}, S.}
\newblock \bibinfo{title}{{Constraining the Speed of Sound inside Neutron Stars with Chiral Effective Field Theory Interactions and Observations}}.
\newblock \textit{ \bibinfo{journal}{\apj}} \textbf{\bibinfo{volume}{860}}, \bibinfo{pages}{149} (\bibinfo{year}{2018}).

\bibitem{Alford:2015dpa}
\bibinfo{author}{Alford, M.~G.}, \bibinfo{author}{Burgio, G.~F.}, \bibinfo{author}{Han, S.}, \bibinfo{author}{Taranto, G.} \& \bibinfo{author}{Zappal{\'a}, D.}
\newblock \bibinfo{title}{{Constraining and applying a generic high-density equation of state}}.
\newblock \textit{ \bibinfo{journal}{Phys. Rev.}} \textbf{\bibinfo{volume}{D92}}, \bibinfo{pages}{083002} (\bibinfo{year}{2015}).

\bibitem{Woosley:2002zz}
\bibinfo{author}{Woosley, S.}, \bibinfo{author}{Heger, A.} \& \bibinfo{author}{Weaver, T.}
\newblock \bibinfo{title}{{The evolution and explosion of massive stars}}.
\newblock \textit{ \bibinfo{journal}{Rev.Mod.Phys.}} \textbf{\bibinfo{volume}{74}}, \bibinfo{pages}{1015-1071} (\bibinfo{year}{2002}).

\bibitem{Umeda:2007wk}
\bibinfo{author}{Umeda, H.} \& \bibinfo{author}{Nomoto, K.}
\newblock \bibinfo{title}{{How much Ni-56 can be produced in Core-Collapse Supernovae? Evolution and Explosions of 30--100 M$_\odot$ Stars}}.
\newblock \textit{ \bibinfo{journal}{Astrophys.J.}} \textbf{\bibinfo{volume}{673}}, \bibinfo{pages}{1014} (\bibinfo{year}{2008}).

\end{thebibliography}
\end{document}